\documentclass{pasa}%
\usepackage{graphicx}
\usepackage{subcaption}
\usepackage{threeparttable}
\usepackage[usenames,dvipsnames]{color}
\title[The MeerKAT Telescope as a Pulsar Facility: MeerTime]{The MeerKAT Telescope as a Pulsar Facility: System verification and early science results from MeerTime}
\author[Bailes, M. et al.]{
M.~Bailes$^{1,2}$\thanks{mbailes@swin.edu.au}, 
A. Jameson$^{1,2}$,
F.~Abbate$^{3,4,5}$, 
E.~D.~Barr$^{4}$,
N.~D.~R.~Bhat$^{6}$,
L.~Bondonneau$^{7,8}$,
M.~Burgay$^{3}$,
S.~J.~Buchner$^9$,
F.~Camilo$^9$,
D.~J.~Champion$^{4}$,
I.~Cognard$^{7,8}$,
P. B.~Demorest$^{10}$,
P.~C.~C.~Freire$^{4}$,
T. Gautam$^{4}$,
M.~Geyer$^9$,
J.-M.~Griessmeier$^{7,8}$,
L.~Guillemot$^{7,8}$
H.~Hu$^{4}$,
F.~Jankowski$^{11}$,
S.~Johnston$^{12}$,
A.~Karastergiou$^{13}$,
R.~Karuppusamy$^{4}$,
D.~Kaur$^{6}$,
M.~J.~Keith$^{11}$, 
M.~Kramer$^{4}$,
J.~van Leeuwen$^{14, 15}$,
M.~E.~Lower$^{1,12}$, 
Y.~Maan$^{14}$, 
M.~A.~McLaughlin$^{16}$, 
B.~W.~Meyers$^{17,6}$, 
S.~Os{\l}owski$^1$, 
L.~S.~Oswald$^{13}$,
A.~Parthasarathy$^{1,2,4}$,
T.~Pennucci$^{10,18}$, 
B.~Posselt$^{13,19}$,
A.~Possenti$^{3,20}$,
S.~M.~Ransom$^{10}$,  
D.~J.~Reardon$^1$,  
A.~Ridolfi$^{3,4}$,
C.~T.~G.~Schollar$^9$,
M.~Serylak$^{9,21}$, 
G.~Shaifullah$^{14}$,
M.~Shamohammadi$^{1,2}$,
R.\,M.~Shannon$^{1,2}$, 
C.~Sobey$^{22}$, 
X.~Song$^{11}$,
R.~Spiewak$^{1,2}$, 
I.~H.~Stairs$^{17}$, 
B.~W.~Stappers$^{11}$,
W.~van Straten$^{23}$,
A.~Szary$^{14,24}$,
G.~Theureau$^{7,8}$,
V.~Venkatraman~Krishnan$^{4}$, 
P.~Weltevrede$^{11}$,
N.~Wex$^{4}$,
T.~D.~Abbott$^9$, 
G.~B.~Adams$^9$,
J.~P.~Burger$^9$,
R.~R.~G.~Gamatham$^9$,
M.~Gouws$^9$,
D.~M.~Horn$^9$,
B.~Hugo$^{9,25}$,
A.~F.~Joubert$^9$,
J.~R.~Manley$^9$,
K.~McAlpine$^9$,
S.~S.~Passmoor$^9$,
A.~Peens-Hough$^9$,
Z.~R.~Ramudzuli$^9$,
A.~Rust$^9$,
S.~Salie$^{9}$,
L.~C.~Schwardt$^{9}$,
R.~Siebrits$^{9}$,
G.~Van~Tonder$^{9}$,
V.~Van~Tonder$^{9}$,
M.~G.~Welz$^{9}$
\\

\affil{$^1$ Centre for Astrophysics and Supercomputing, Swinburne University of Technology, P.O. Box 218, Hawthorn, VIC 3122, Australia} 
\affil{$^2$ ARC Centre of Excellence for Gravitational Wave Discovery (OzGrav)}
\affil{$^3$ INAF - Osservatorio Astronomico di Cagliari, Via della Scienza 5, 09047 Selargius (CA), Italy}
\affil{$^{4}$ Max-Planck-Institut f\"{u}r Radioastronomie, Auf dem H\"{u}gel 69, D-53121 Bonn, Germany}
\affil{$^{5}$ Dipartimento di Fisica `G. Occhialini', Universit\`a degli Studi Milano - Bicocca, Piazza della Scienza 3, 20126, Milano, Italy}
\affil{$^{6}$ International Centre for Radio Astronomy Research (ICRAR), Curtin University, 1 Turner Avenue, Technology Park, Bentley, WA 6102, Australia}
\affil{$^{7}$ Laboratoire de Physique et Chimie de l'Environnement et de l'Espace LPC2E CNRS-Universit{\'e} d'Orl{\'e}ans, F-45071 Orl{\'e}ans, France}
\affil{$^{8}$ Station de radioastronomie de Nan{\c c}ay, Observatoire de Paris, PSL Research University, CNRS/INSU F-18330 Nan{\c c}ay, France}
\affil{$^{9}$ South African Radio Astronomy Observatory, 2 Fir Street, Black River Park, Observatory 7925, South Africa}
\affil{$^{10}$ National Radio Astronomy Observatory, 520 Edgemont Rd., Charlottesville, VA 22903, USA}
\affil{$^{11}$ Jodrell Bank Centre for Astrophysics, Department of Physics \& Astronomy, The University of Manchester, Manchester M13 9PL, UK}
\affil{$^{12}$ CSIRO Astronomy \& Space Science, Australia Telescope National Facility, P.O. Box 76, Epping, NSW 1710, Australia} 
\affil{$^{13}$ Department of Astrophysics, University of Oxford, Denys Wilkinson Building, Keble Road, Oxford OX1 3RH, UK}
\affil{$^{14}$ ASTRON, The Netherlands Institute for Radio Astronomy, Postbus 2, NL-7900 AA Dwingeloo, the Netherlands}
\affil{$^{15}$ Anton Pannekoek Institute for Astronomy, University of Amsterdam, Science Park 904, 1098 XH Amsterdam, Netherlands}
\affil{$^{16}$ Department of Physics and Astronomy, West Virginia University, Morgantown, WV 26506-6315 and the Center for Gravitational Waves and Cosmology, Morgantown, WV 26505}
\affil{$^{17}$ Department of Physics and Astronomy, University of British Columbia, 6224 Agricultural Road, Vancouver, BC V6T 1Z1, Canada}
\affil{$^{18}$ Institute of Physics, E\"{o}tv\"{o}s Lor\'{a}nd University, P\'{a}zm\'{a}ny P. s. 1/A, 1117 Budapest, Hungary}
\affil{$^{19}$ Department of Astronomy \& Astrophysics, Pennsylvania State University, 525 Davey Lab, 16802 University Park, PA, USA}
\affil{$^{20}$ Universit\'a di Cagliari, Dip di Fisica, S.P. Monserrato-Sestu Km 0,700 - 09042 Monserrato (CA), Italy}
\affil{$^{21}$ Department of Physics and Astronomy, University of the Western Cape, Bellville, Cape Town 7535, South Africa}
\affil{$^{22}$ CSIRO Astronomy and Space Science, 26 Dick Perry Avenue, Kensington, WA 6151, Australia}
\affil{$^{23}$ Institute for Radio Astronomy \& Space Research, Auckland University of Technology, Private Bag 92006, Auckland 1142, NZ}
\affil{$^{24}$ Janusz Gil Institute of Astronomy, University of Zielona G\'ora, Lubuska 2, 65-265 Zielona G\'ora, Poland}
\affil{$^{25}$ Department of Physics and Electronics, Rhodes University, PO Box 94, Grahamstown 6140, South Africa}

}
\jid{PASA}
\doi{10.1017/pas.\the\year.xxx}
\jyear{\the\year}
\usepackage{aas_macros}
\usepackage{hyperref}
\begin{document}
\begin{frontmatter}
\maketitle
\begin{abstract}
We describe system verification tests and early science results from the pulsar processor (PTUSE) developed for the newly-commissioned 64-dish SARAO MeerKAT radio telescope in South Africa.
MeerKAT is a high-gain ($\sim$\,2.8\,K/Jy) low-system temperature ($\sim$\,18\,K at 20cm) radio array that currently operates from 580--1670\,MHz and can produce tied-array beams suitable for pulsar observations. 
This paper presents results from the MeerTime Large Survey Project and commissioning tests with PTUSE. Highlights include observations of the double pulsar J0737$-$3039A, pulse profiles from 34 millisecond pulsars from a single 2.5\,h observation of the Globular cluster Terzan 5, the rotation measure of Ter5O, a 420-sigma giant pulse from the Large Magellanic Cloud pulsar PSR~J0540$-$6919, and nulling identified in the slow pulsar PSR~J0633--2015. One of the key design specifications for MeerKAT was absolute timing errors of less than 5\,ns using their novel precise time system. Our timing of two bright millisecond pulsars confirm that MeerKAT delivers exceptional timing. PSR~J2241$-$5236 exhibits a jitter limit of $<$\,4\,ns per hour whilst timing of PSR~J1909$-$3744
over almost 11 months yields an rms residual of 66\,ns with only 4\,min integrations.
Our results confirm that the MeerKAT is an exceptional pulsar telescope. The array can be split into four separate sub-arrays to time over 1000 pulsars per day and the future deployment of S-band (1750--3500\,MHz) receivers will further enhance its capabilities.
\end{abstract}
\end{frontmatter}

\section{INTRODUCTION}
\label{sec:intro}

In the standard model, radio pulsars are highly-magnetised rapidly rotating neutron stars that emit a coherent light-house beam of often highly polarised radio emission directed by
their magnetospheres \citep{2004hpa..book.....L}. 
The weak braking torques caused by their rapidly rotating magnetic fields and their high moments of inertia make them extremely stable flywheels, and it is often possible to predict the pulsar spin period and indeed pulse phase years in advance of observation \citep{1992RSPTA.341..117T}. 
Most radio pulsars regularly emit irregular single-pulse shapes that usually sum to an average profile within 1000 rotations that is often remarkably constant \citep{2012MNRAS.420..361L}.
Timing of these mean profiles against a template produces an arrival time which can be used to derive a model of the pulsar's spin-down, astrometric and binary parameters and 
propagation through the ionised interstellar medium. 
The frequency-dependence of the pulse arrival time is well described by the cold plasma dispersion relation, and allows observers to compute the column density of free electrons along the line of sight to the observer. 
The integral of this column density is referred to as the pulsar's dispersion measure (DM). 
The most accurate pulse arrival times require observers to remove the broadening
of the pulse profile across the finite channel bandwidths using a process known as
coherent dedispersion \citep{1975MComP..14...55H} and accurately monitor changes 
in the DM \citep{2013MNRAS.429.2161K}.

According to version 1.62 (Feb 2020) of the Australian Telescope National Facility pulsar catalogue\footnote{https://www.atnf.csiro.au/research/pulsar/psrcat/}
\citep{mhth05} there are currently 2800 pulsars known, $\sim 97$\% of which are visible at radio wavelengths. 
Radio pulsars range in pulse period ($P$) from 1.4\,ms to 23.5\,s, and have inferred dipolar magnetic field strengths from 5$\times10^7$G to $\sim$10$^{15}$\,G. Over 10 percent of known pulsars are members of binary systems, and the majority of these are the so-called `recycled pulsars', that have had their magnetic fields weakened and spin periods shortened by mass accretion from a donor
\citep{1991PhR...203....1B}. The fastest pulsars ($P<20$\,ms) are usually referred to as `millisecond pulsars' (MSPs).
These pulsars are often in very clean systems well approximated by point masses and are ideal for tests of gravitational and stellar evolution theories \citep{2016ApJ...829...55W}.

State of the art pulsar timing allows us to measure pulse arrival times to better than one part in 10$^4$ of the pulse period \citep{2001Natur.412..158V}, leading to sub-microsecond arrival times for the MSPs. In their most recent data release (dr2) the International Pulsar Timing Array\footnote{www.ipta4gw.org} lists an rms timing residual for the bright 5.7\,ms MSP PSR~J0437$-$4715 of just 110\,ns and 14 others with residuals below 1\,$\mu$s \citep{2019MNRAS.490.4666P}.

Modern radio telescopes can detect radio pulsars 
with a mean flux density (ie averaged over the pulse period)
down to just a few $\mu$Jy in very deep pointings, and the large-scale surveys of much of the galactic plane are complete to $\sim$0.1 mJy \citep{2015MNRAS.450.2922N}. 
The population exhibits a standard log-$N$/log-$S$ distribution consistent with a largely planar distribution with a slope of $\sim-$1.  The most compelling pulsar science is usually derived from accurate pulse timing which for most pulsars is signal-to-noise limited as the vast majority of known pulsars have flux densities less than 1\,mJy at 1400 MHz. For this reason the field has been dominated by the world's largest radio telescopes that possess low-temperature receivers and digital backends capable of coherently dedispersing the voltages induced in the receiver by the radio pulsars. 
These telescopes can produce the high signal-to-noise profiles required to 
test theories of relativistic gravity, determine neutron star masses, clarify the poorly-understood radio emission mechanism, and relate the latter to the magnetic field topology.

The galactic centre is at declination $\delta=-29^\circ$ and this makes the Southern hemisphere a particularly inviting location for pulsar studies. For many years the Parkes 64\,m telescope has had almost exclusive access to radio pulsars south of declination $\delta=-$35$^\circ$, and 
consequently discovered the bulk of the pulsar population.  
When choosing a site and host country for the forthcoming Square Kilometre Array \cite[][]{2009IEEEP..97.1482D}  SKA1-mid telescope, the strong pulsar science case made Southern hemisphere locations particularly 
desirable. MeerKAT \citep{jonas2009}  is the South African SKA precursor telescope located at the future site of SKA1-mid and the full array has four times the gain (i.e.~2.8\,K/Jy) of the Parkes telescope (0.7\,K/Jy). 
The first receivers (L-band) to come online possess an excellent system temperature ($\sim$18\,K) along with 856\,MHz of recorded bandwidth. The pulsar processor often just records the inner 776\,MHz of this for science purposes.
The telescope is located at latitude $-30^{\circ}43'$ and is ideal for studies of the large population of southern pulsars and those in the Large and Small Magellanic Clouds. Much of pulsar science is signal-to-noise limited until a pulsar hits its `jitter limit' (the lowest timing residual
obtainable due to pulse-to-pulse variability in the individual pulses - see \citep{2014MNRAS.443.1463S} ). When far from the jitter limit 
the timing error is inversely proportional to the signal to noise ratio. 
For most pulsars in the Parkes Pulsar Timing Array \citep{2013PASA...30...17M}, the limit is rarely reached when observed with the Parkes 64\,m telescope
unless the pulsar is experiencing a scintillation maximum \citep{sod+14}. 
A notable exception is
the bright MSP PSR~J0437$-$4715, that is always jitter-limited when observed at the Parkes telescope \citep{2011MNRAS.418.1258O} due to its large 1400\,MHz mean flux density of 150\,mJy \citep{2015MNRAS.449.3223D}. As telescopes become more sensitive, the number of pulsars in the same
integration time being jitter-limited increases.

The South African Radio Astronomy Observatory (SARAO) owns and operates MeerKAT and, 
before it was commissioned, called for Large Survey Projects (LSPs) that could exploit the telescope's scientific potential. The MeerTime\footnote {http://www.meertime.org} \citep{2018arXiv180307424B} collaboration was successful at obtaining LSP status and commenced its first survey observations in February of 2019. This paper reports on MeerTime's validation of MeerKAT as a pulsar telescope and presents some early science results from its four major themes: Relativistic and Binary Pulsars, the Thousand Pulsar Array \citep{johnstonetal2020}, Globular Clusters and the MeerKAT Pulsar Timing Array.

A glimpse of MeerKAT's potential as a pulsar telescope was presented in \citep{2018ApJ...856..180C} when it was part of a campaign that observed the revival of the magnetar PSR~J1622$-$4950. Since then there have been a number of developments of the system that enable a wider range of 
pulsar observing modes that will be discussed forthwith.

The structure of this paper is as follows: In section \ref{sec:MKP}, we provide an overview of MeerKAT as a pulsar telescope including examples of the UHF and L-band radio bands, 
the Precise Time Manager (PTM), choice of polyphase filterbanks, and the SKA1 prototype pulsar processor PTUSE developed by Swinburne University of Technology. In section \ref{sec:SV}, we describe our validation of the system and pulsar hardware before presenting new science from observations of selected pulsars and globular clusters in section \ref{sec:RESULTS}. Finally, we briefly discuss some of the prospects for the future of this facility including new modes, receivers, and extensions and ultimate extension to become SKA1-mid in section \ref{sec:discussion}.

\section{The MeerKAT telescope as a Pulsar Facility}
\label{sec:MKP}

\begin{figure*}
    \centering
    \includegraphics[scale=0.7]{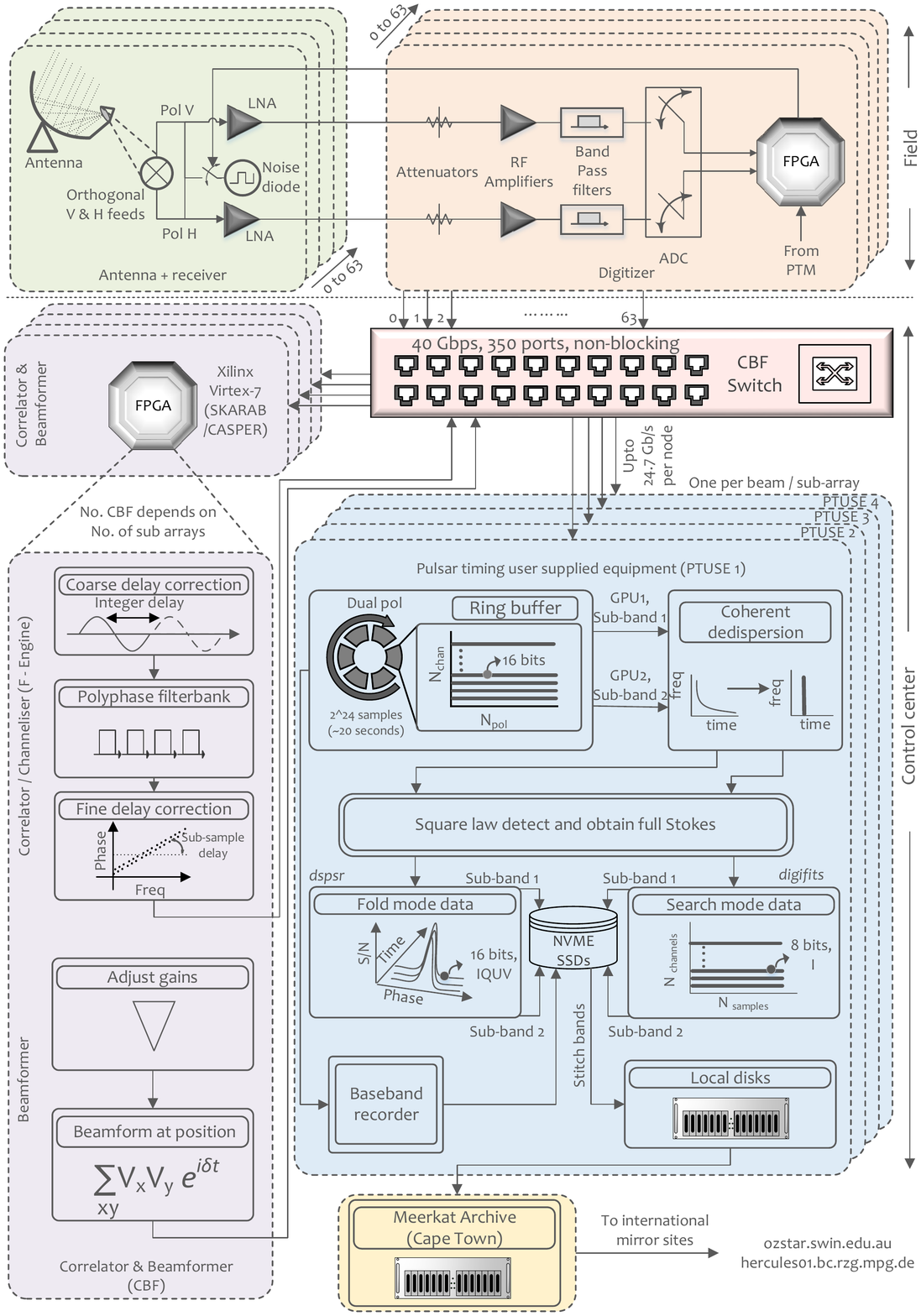}
    \caption{ A block diagram of the signal chain for MeerTIME observations. Signals from all antennas are digitised in the field and sent to the correlator-beamformer (CBF) engine in the building via the CBF switch. For pulsar observations, CBF performs channelisation (1K or 4K mode) and beamforms at the requested sky position. The beamformed voltages are sent to the PTUSE machines via the CBF switch at a data rate of up to 24.7 Gb/s. Each PTUSE node can process one beam/sub-array. In each PTUSE node, the incoming voltages are temporarily stored in a ring buffer from which they are sent to the two GPUs, each processing one half of the band. The GPUs perform coherent dedispersion and square law detection to obtain 16-bit, full Stokes data, which, depending on the observation, is either folded into pulsar archives or scrunched into 8-bit, total intensity filterbank data. The GPUs write the end products to the NVME disks via the CPUs, from which the two bands are stitched together and transferred to the local disks. The baseband recorder, when triggered, can dump baseband data to the NVME disks for a period of about 40 minutes. The data from the local disks are eventually transferred to the MeerKAT archive in Cape Town, from which they are transferred to international mirror sites. }
    \label{fig:block_diagram}
\end{figure*}

A high level block diagram of the system is provided in Figure \ref{fig:block_diagram} that
describes the system all the way from the antennas to the final data product archive.
The backend system design largely followed the CASPER philosophy of transferring
as much of the transport between the digital subsystems to commodity-off-the-shelf 
(COTS) components and industry standard protocols (eg ethernet) that involve commercial switches and is interfaced to the pulsar
processor which is itself a modern server comprised entirely of COTS components.

\subsection{The MeerKAT Radio Frequency Spectrum}
\label{RFS}

MeerKAT is located in the Karoo, some 450 km north-east of Cape Town in the Northern Cape Province. Its low population density makes it an attractive site to pursue radio astronomy.  The low-frequency HERA experiment \citep{2017PASP..129d5001D} and the future 197-dish SKA1-mid telescope \citep{2009IEEEP..97.1482D} -- of which MeerKAT will be a part -- will be located at the site which is protected by legislation against 
radio transmissions in many bands of relevance to radio astronomers. 

The technology behind low-noise amplifiers and radio receivers has greatly improved since the dawn of radio astronomy in the 1950s. 
Whilst even just a couple of decades ago it was necessary to sacrifice fractional bandwidth to minimise system temperature new engineering practices and technologies now permit the development of low-noise ($\sim$20\,K) receivers over a full octave or more of bandwidth eg. \citep{2019arXiv191100656H}.

The original MeerKAT radio telescope specification had a target effective collecting area per unit receiver temperature of $A_{\rm eff}/T$ = 220 m$^2$/K but remarkably achieved 350--450~m$^2$/K (depending upon radio frequency), well over a factor of 2 increase in observing efficiency over the design specification. These figures equate to a system equivalent flux density (\textrm{SEFD}) of $\textrm{SEFD}=T/G\sim7$\,Jy, 
where $T$ is the system temperature in K and $G$ is the total antenna gain $G= A{\eta}/(2k)$,  $A$\, is the collecting area, $\eta$ is the  aperture efficiency  and $k$ is Boltzmann's constant. The total MeerKAT antenna gain is 2.8\,K/Jy and the system temperature about 18\,K in the optimal 
location of the 1400\,MHz band. The receivers have two orthogonal linear polarisations (H and V for horizontal and vertical respectively). 

\begin{figure*}
    \centering
    \begin{subfigure}{0.45\textwidth}
        \includegraphics[angle=0,width=\columnwidth]{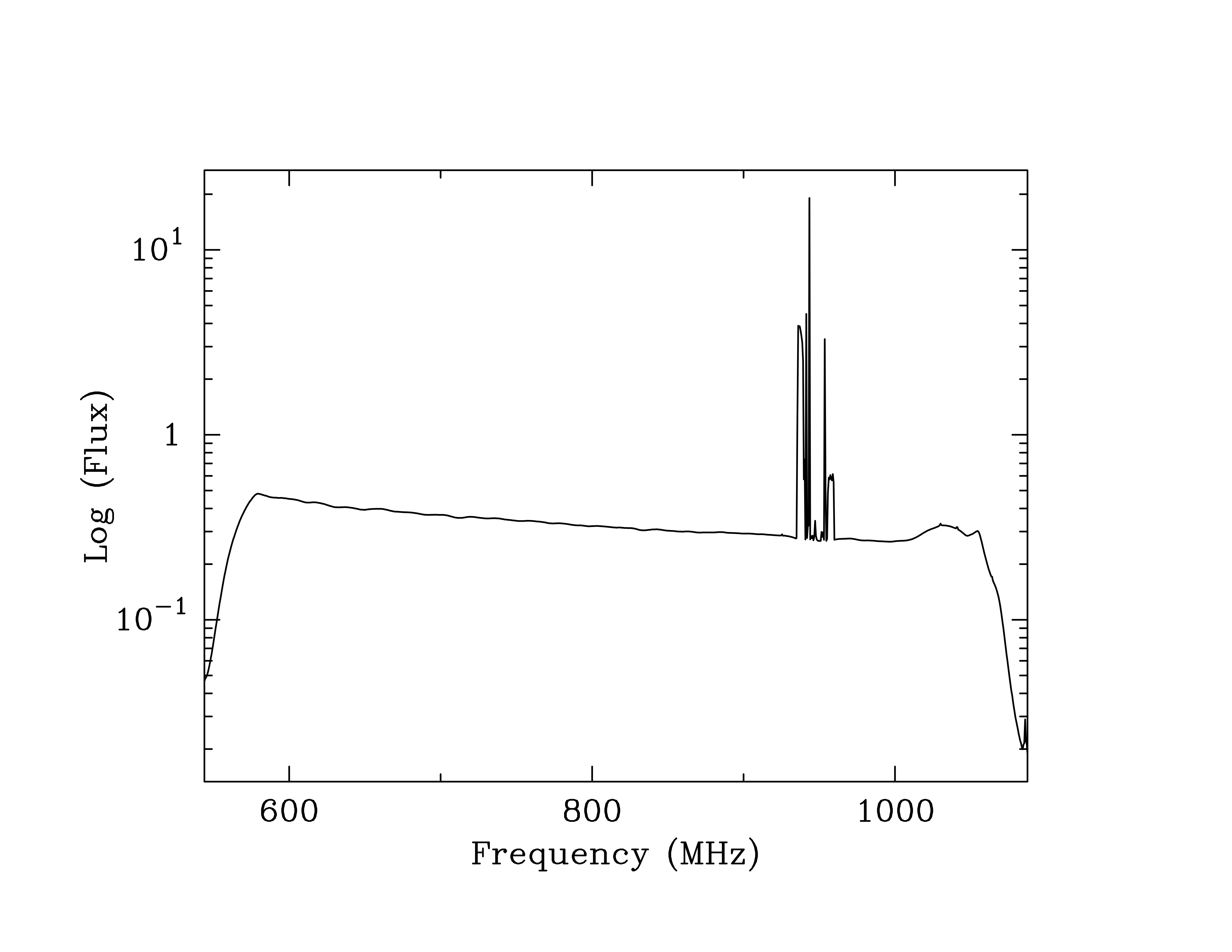}
        \caption{UHF receiver bandpass for Stokes I (544--1088\,MHz).}
        \label{fig:UHF_bandpass}
    \end{subfigure}
    \hfill
       \begin{subfigure}{0.45\textwidth}
        \includegraphics[angle=0,width=\columnwidth]{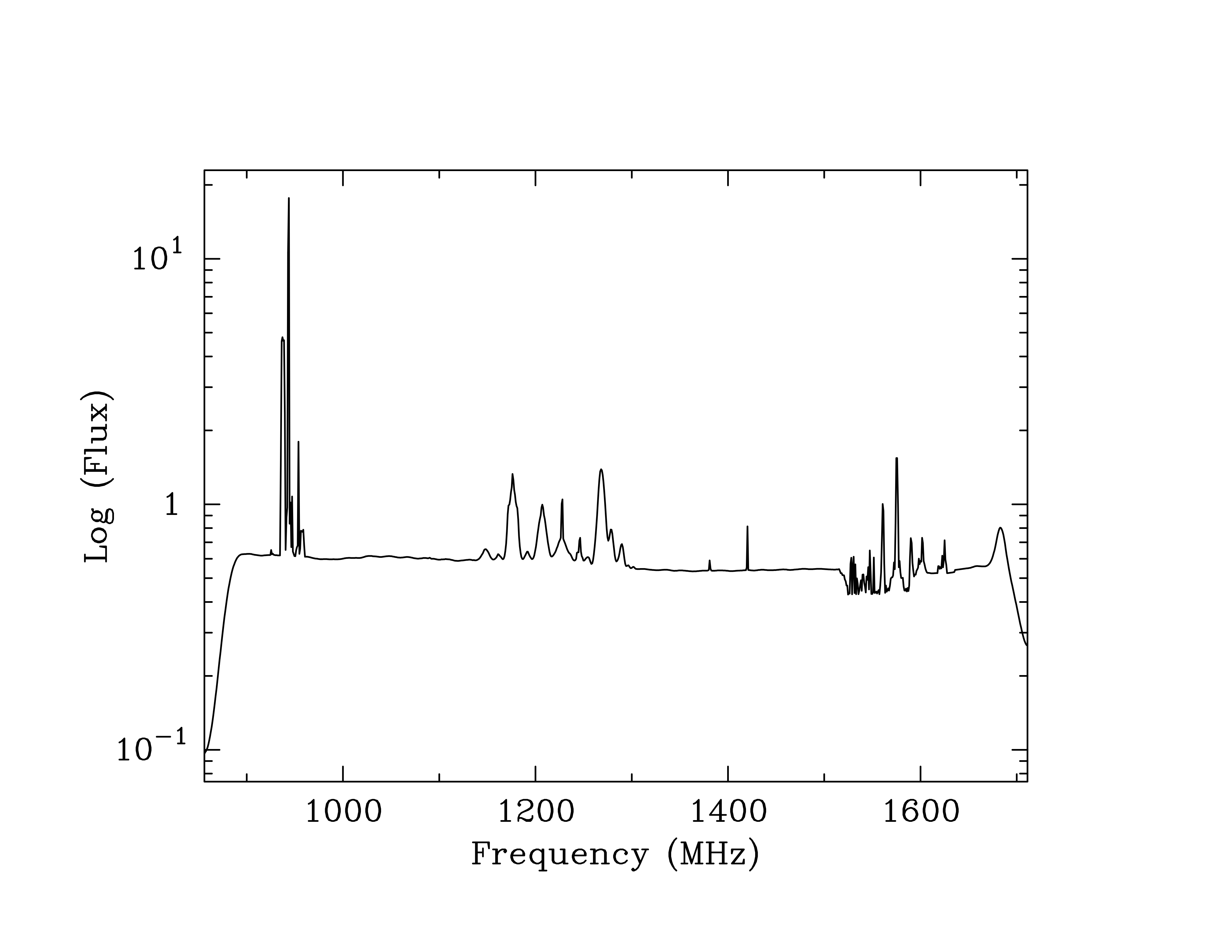}
        \caption{L-Band receiver bandpass for Stokes I (856--1712\,MHz)}
        \label{fig:Lband_bandpass}
    \end{subfigure}
    \caption{The post-calibration bandpasses of a tied-array beam, for MeerKAT's UHF and L-Band receivers. The flux density scale is arbitrary. It is often difficult
    to completely flatten the band in regions of persistent interference such as that
    near 1530-1600 MHz.}
    \label{fig:bandpasses}
\end{figure*}

In Figure \ref{fig:bandpasses} we present the radio spectrum as observed from the MeerKAT site for the UHF (544--1088\,MHz) and L-band (856--1712\,MHz) receivers taken from observations
of the double pulsar PSR~J0737$-$3039A. 
Pulsar observations are usually made with 1024 or 4096 frequency channels.
The UHF band is remarkably clean, with just some small (strong but narrow-bandwidth) residual mobile phone-related transmissions visible around 940 MHz. In most countries that house large-diameter (64m class and above) telescopes, the UHF band is so badly polluted by digital television and mobile phone transmissions that it is often unusable except in very narrow frequency windows some 10s of MHz wide. Both of the SKA sites appear to have been chosen well and offer a renewed opportunity to explore the Universe
at these frequencies.
For pulsars, this is especially relevant, as most possess steep spectra \citep{1998ApJ...506..863T, 2000A&AS..147..195M, 2018MNRAS.473.4436J}, with spectral indices of between $-1$ and $-3$ above 1 GHz.

The 1400-MHz (L-band) receiver band is not as pristine as the UHF band, but still has much of the spectrum available for science (see a quantified
analysis below), depending upon the flux density of the target pulsar. The tied-array
beam helps dilute interfering signals by dephasing them but the large number of bits in the digitizers and beam-former that deliver accurate channelisation of the data have one drawback in the sense that the interference-to-noise ratio can
be extremely high. This makes deletion of at least some frequency
channels essential before integration across frequency channels.

Like all modern observatories, the 1400\,MHz band suffers from 
transmissions from Global Navigation Satellite Systems (GNSS) and other satellites that are extremely strong and impossible to avoid. The small apertures (and hence large side-lobes) that the 14-m dishes of MeerKAT provide make satellite transmissions in the band almost omnipresent. 
The L-band spectrum is shown in Fig \ref{fig:Lband_bandpass}. 

The first of the S-band ($1.75-3.5$\,GHz) receivers of the Max-Planck-Institut f\"ur Radioastronomie \citep{2016mks..confE...3K} are currently being installed and tested. When fully installed these will provide the possibility of performing high precision timing at even higher frequencies.

\subsection{Precise Time Systems in MeerKAT}
\subsubsection{Background on requirements}
\label{PTM}
The SKA phase I (comprising SKA1-low and SKA1-mid) has been strongly motivated by two key science projects, the Epoch of Reionisation and strong-field tests of gravity respectively \citep[e.g.,][]{2016mks..confE...3K}.
Despite the advent of direct gravitational wave detection \citep{Abbott+16} and black hole imaging \citep{EHT+19}, a number of precision strong-field tests can only be achieved at radio wavelengths using pulsars. This includes the detection of a gravitational wave background from supermassive black holes using pulsar timing arrays \citep[e.g. ][]{2015Sci...349.1522S, Lentati+15,  Nanograv11yr}, which will require timing an ensemble of MSPs to precisions well below 100\,ns and possibly down to 10\,ns.
In order to achieve such precisions, two of the SKA1 specifications are especially relevant, one related to the calibration of the polarimetry that otherwise leads to systematic errors in timing; see, e.g., \citep{fkp+15}, and the other the knowledge of absolute time with respect to Coordinated Universal Time (UTC) over a full decade. The SKA1-mid specification on calibratable polarisation purity is $-$40\,db, and on time 5\,ns over 10 years.

The precise time systems in MeerKAT are specified to provide time products accurate to better than 5\,ns, relative to UTC. The telescope was designed for absolute, not just relative timing which is the norm. This is achieved via a first-principles approach, by managing the time delays associated with every element of the geometric and signal paths.

\subsubsection{Realisation of system timing for the MeerKAT telescope}
MeerKAT has defined a reference point on the Earth to which all time is referred. It is a location a few hundred metres approximately north of the centre of the array with with the ITRF coordinates X=5109360.0, Y=2006852.5, Z=$-$3238948.0\,m. $\pm$0.5 m. The position of the reference point was chosen as the circumcentre of the array, roughly a metre above the ground. There is no antenna at this point; all pulsar timing is ultimately referred to when an incident radio wave would have struck this point. This location for MeerKAT is installed in the observatory coordinate file for the pulsar timing software \textsc{tempo2} \citep{Hobbs+06}, and used for all work in this paper.

The coordinates of the source, antenna and UT1 define the geometric delay, and every attempt is made to derive the further delays incurred by the signal as it reflects off the telescope surfaces, passes through the feed/receiver/cable/filter and is ultimately sampled by the analogue to digital converters (ADCs). Round-trip measurements account for cable and fibre delays and are accurate to typically 1\,ns. Estimates of the error in each stage of the path are also recorded for later dissemination to the pulsar processor to be recorded with the data.

\begin{figure}
    \centering
    \includegraphics[scale=0.36]{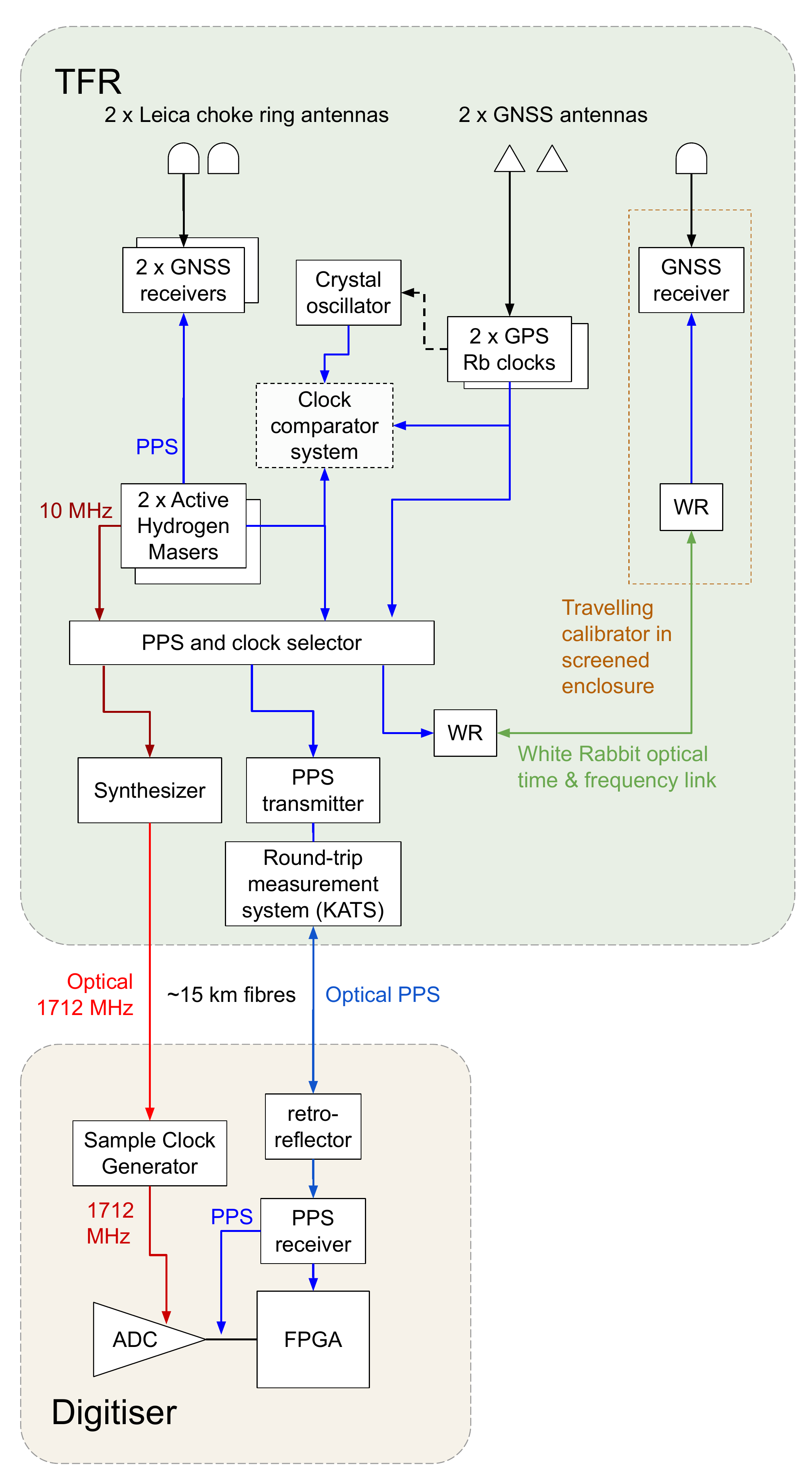}
    \caption{MeerKAT time systems showing time transfer from GNSS, the local clock ensemble and transfer of time to the digitisers. Here Rb stands for the element Rubidium, WR is the White Rabbit sub-ns accuracy Ethernet-based 
    time transfer system. PPS is the 
    1 pulse per second used in precision timing experiments, and FPGAs are Field Programmable Gate Arrays. See text for further details.}
    \label{fig:time_transfer}
\end{figure}

At each antenna, a digitiser is mounted close to the focus (see the lower panel in Figure \ref{fig:time_transfer}). The digitiser ADC is driven by the digitiser sample clock, derived from the Maser frequency standards and disseminated directly to the digitisers by optical fibre. The input voltages are sampled at the Nyquist rate after passing through an analogue filter for the selected band. The L-band receiver digitizes the data at exactly 1712\,MHz (real samples) and passes the second Nyquist zone (i.e., top half of the band 856-1712\,MHz) to the correlator-beamformer via optical fibre Ethernet. 
The digitiser maintains a 48-bit ADC sample counter, used to tag every packet. The counter is reset every day or two by telescope operators, before it overflows. The 10-bit digitiser offers excellent resistance to radio frequency interference and makes it possible to confine RFI to only the relevant frequency channels unless it causes saturation of the ADC. 

Optical timing pulses are generated by the Time and Frequency Reference (TFR) system in the processor building and disseminated to all antennas via dedicated optical fibres. The digitiser records the time of arrival of the optical pulse from the masers; it is also used to reset the digitiser sample clock when necessary. The timing fibres are buried 1\,m deep to minimise diurnal temperature variations, but change length over the year as the site temperature varies. A round-trip measurement system continuously measures their length by timing the round trip of the timing pulse as reflected back from the digitiser \citep{ spie2017,  ifcs2018nr2}. This falls within a general class of time-of-flight measurement which is traceable to the SI unit of time \citep{terra}.

In the correlator, prior to channelisation, the signals from all antennas are buffered and aligned using the ADC sample counter. Integer samples of delay are applied to compensate for physical delays. A multi-tap Hann-window polyphase filterbank is used to filter the data along with a phase gradient to eliminate the remaining (sub-sample)  geometrical delay. Pulsar timing observations at L-band usually use the 1024 channel mode, giving a time resolution of 1/$B$ of ~1.196\,$\mu$s. The narrowest mean MSP profile features are 10s of $\mu$s in width although `giant' pulses have been observed with time resolution from the Crab pulsar with timescales of down to 1\,ns \citep{2003Natur.422..141H}. 

The Precise Time Manager (PTM) is a program that collects and aggregates all known delays in the system: geometric, physical delays in the antenna, analogue and digital delays in the receiver and digitiser, the correlator delay and the digitiser clock offset. PTM computes the \textit{time that the wavefront corresponding to a certain ADC sample crossed the array phase centre}, and the uncertainty in this time. This time is passed to PTUSE for recording in the header of the pulsar observation. In September 2019, the uncertainty in this value was 3 to 4\,ns.

\subsubsection{Tracking of telescope time: Karoo telescope time}
The station clock is referred to as KTT (Karoo Telescope Time). This timescale is generated by the TFR subsystem using an ensemble of two active hydrogen masers, two Rubidium clocks and a quartz crystal. It is a physical timescale, defined at a connector in the system. The masers drift by typically just a few ns per day; KTT is kept $\sim$1\,$\mu$s from UTC by adjusting their synthesized frequency every few months to keep them in defined offset bands. 

The TFR system provides a 10\,MHz frequency reference, from the Maser currently in use. Fractional frequency offset is kept to smaller than $2\times10^{-13}$ with respect to UTC; no significant drift in time occurs during any observing campaign. Frequency synthesizers in the building, one per band, are locked to the reference frequency and generate the 1712\,MHz and 1088\,MHz sample clocks for distribution to the digitisers. 

The TFR also provides accurate time to other components of the telescope via Precision Time Protocol (PTP): this is used amongst other things in the pointing of the telescope and control of the precision timing systems \citep{ifcs2018nr2}. 

Pulsar timing requires the difference between UTC and KTT to be measured. At MeerKAT, this is done with a set of calculations to calculate ensemble time using interclock differences between five clocks, and measurements of four clocks with respect to GPS via two dual-band GNSS receivers and two single band GPS-only receivers \citep{spieKTT}. The multiple measurements then lead to clock solutions for each of the clocks in the ensemble via linear combinations of the different measurements. The usage of multiple clocks enables error/instability detection in any one of the clocks and a lower variance estimate in any one of the clocks as with standard ensembling used in timescale generation \citep{levine}.

Reference is made to the UTC via common-view comparison with the National Metrology Institute of South Africa (NMISA) in Pretoria, and by direct comparison with UTC(USNO) via GPS time dissemination. The uncertainty of the absolute time difference between KTT and UTC is specified to be less than 5\,ns. At present, the systematic (non-varying) offset is only \emph{stated} to 50\,ns due to verification of absolute offset calibration being undertaken. The offset calibration was performed using absolutely calibrated GPS receivers before main observations were being undertaken, and will in future be done using an EMC-quiet calibrator \citep{ifcs2018nr1} in order not to disrupt the observations. The repeatability between observations is thought to be about 5\,ns, implying that the systematics and the stability will finally converge to the latter number; final absolute calibration via a GPS simulator traceability chain will lower the absolute offset to <1\,ns. As we shall see later on, there is evidence from our pulsar timing results that we are approaching these levels of clock correction/stability. Furthermore, the clock tracker is currently being improved from a semi-real time predictive method, to a fully post-facto non-causal filtering type, using Savitzky-Golay filtering \citep{sg64} on which provisional internal self-consistency checks suggest a numerical error of <1\,ns. Post-facto \citep{levine}, non-causal calculation is always better in improving timing compared to real-time `UTC-like' timescale estimation, due to an increased data set, administrative oversight, ability to correct for non-idealities and the inherent outperformance of smoothers as compared to causal filters \citep{Einecke, jensen2012}.

\subsection{The Polyphase Filterbanks}
\label{PFB}
The MeerKAT channeliser (the F-engine) uses a polyphase filterbank (PFB) to channelise the digitized bandwidth into 1024 or 4096 critically sampled frequency channels with configurations described in Table \ref{tb:fengine-configuration}. A PFB filter design with a 16-tap Hann window was deployed aiming to achieve high sensitivity for
continuum mapping with minimal bandwidth losses. 
This design uses only 6dB of attenuation at the channel edges which however gives rise to significant aliasing from adjacent channels in pulsar observations.
To address this, an alternate 16-tap Hann window design that provides superior spectral purity performance at a modest price in sensitivity (due to reduced effective bandwidth) was implemented. This was achieved by reducing the 6\,dB cut-off frequency to 0.91 times the channel width, sacrificing $\sim$5\% of the sensitivity to reduce the leakage by 10\,dB.
Delay compensation is done in the F-engine. The requested delay polynomial is re-computed at every FFT. Coarse delay is done with a whole-sample delay buffer before the PFB. Fine delay is applied by phase rotation after the PFB. The channelised voltages in each packet from the F engine are time-stamped with the same 48-bit counter as the original digitiser voltages, delayed by the delay tracking system, and the impulse response of the channeliser. 
The F-engine operates internally with 22-bit complex numbers, these are requantised to 8 bits real + 8 bits imaginary for transmission over the network. The requantisation gain is chosen to provide adequate resolution on the quietest channels, which results in some of the strongest RFI channels being clipped. The requantisation gain register is also used to flatten the bandpass, equalise the gain of H and V polarisations, and correct for per-channel phase variations found during phase-up. 

\begin{figure}
    \centering
 	\includegraphics[angle=0,width=\columnwidth]{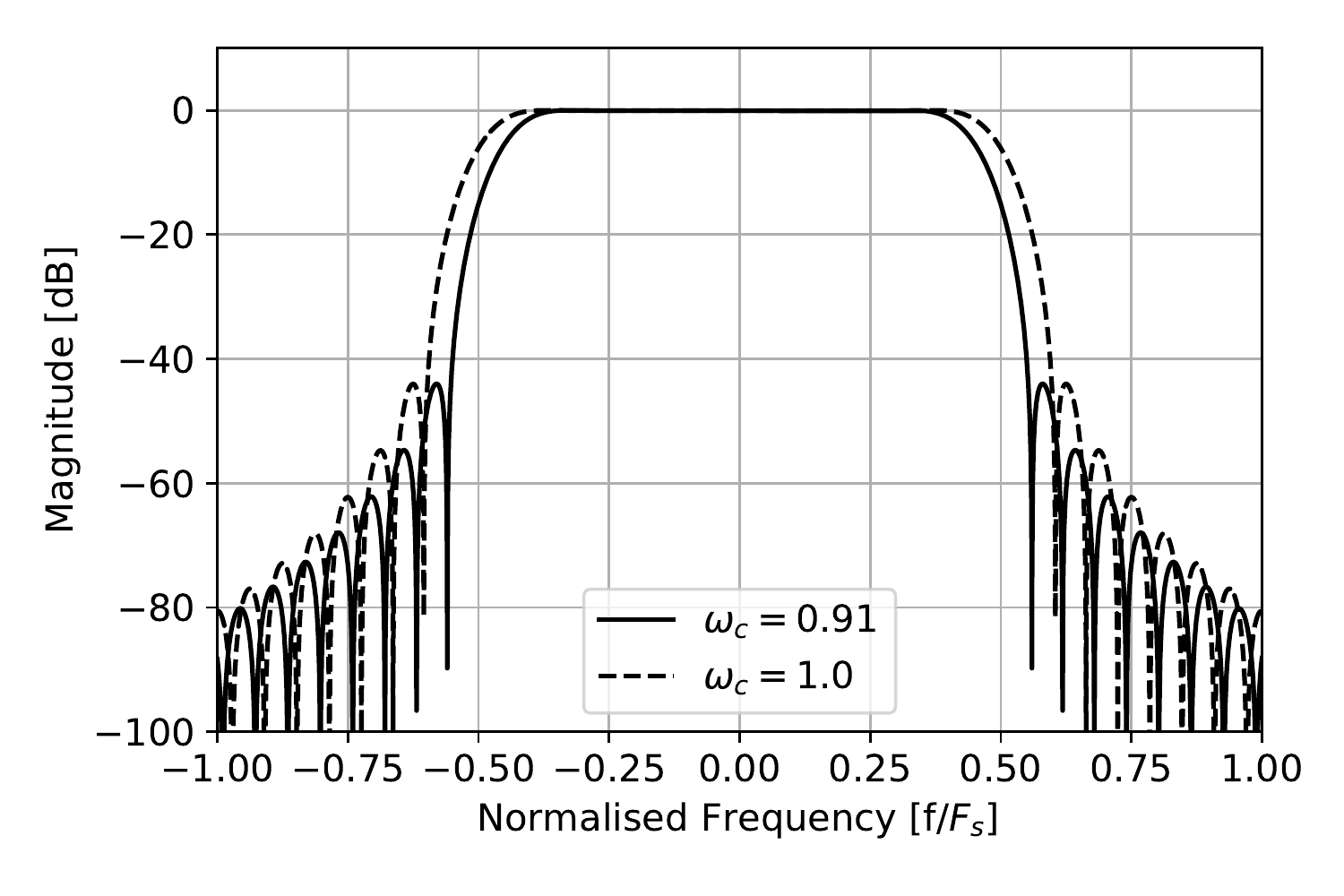}
 	\caption{Magnitude response of the two evaluated channeliser filter designs.
 	The original filter design (dashed) led to significant artifacts in pulsars with
 	high ratios of dispersion measure to period - see text.}
 	\label{fig:hann_fir_response}
\end{figure}

A comparison of the shape of the transfer function for the two modes is shown in Figure \ref{fig:hann_fir_response}. The level magnitude response at channel boundaries (-0.5 and +0.5) determine the level of spectral leakage artifacts in pulsar timing observations. This effect is discussed in section \ref{leakage}, which shows the 0.91 filter design greatly reduces the artifacts and hence why this design has been deployed for most MeerTime observations (exceptions are those made with the 4096 channel mode).

\begin{table*}
    \caption{MeerKAT F-Engine Configurations, with each producing dual polarisations quantised to 8 bits per sample. Frequencies and bandwidths are quoted in MHz and the sampling interval in microseconds.}
    \label{tb:fengine-configuration}
    \begin{threeparttable}
    \begin{tabular}{llllll}
        \hline
        Band & Approx. Centre Frequency & Bandwidth & Channels & Sampling Interval & Data Rate \\
         &  (MHz) $\dagger$ & (MHz) & ($N_{\rm chan}$) & ($\mu$s) & (Gbits/s) \\
        \hline
        L & 1284 & 856 & 1024 & 1024/856 & 27.392 \\
        L & 1284 & 856 & 4096 & 4096/856 & 27.392 \\
        UHF & 816 & 544 & 1024 & 1024/544 & 17.408 \\
        UHF & 816 & 544 & 4096 & 4096/544 & 17.408 \\
        \hline \\
    \end{tabular}
    \begin{tablenotes} 
    \item[$\dagger$] The F-engine PFB implementation lowers the precise centre frequency of all channels  by half a fine channel width (i.e. by $BW / N_{\rm chan} / 2$), where $BW$\, is the total bandwidth. 
    For example, the precise centre frequency of the first channel for L band with 1024 channels is 856\,MHz.
    \end{tablenotes}
    \end{threeparttable}
\end{table*}

\subsection{The Beamformer}
\label{Bengine}
The MeerKAT beamformer (the B-engine) creates a dual polarisation tied-array beam by adding together the channelised complex voltages for all antennas, as produced by their individual F-engines. Thus the beam is also Nyquist-sampled like the antenna voltages. The B-engine is distributed among (typically) 64 SKARABs (custom boards developed by SARAO for digital signal processing designed to be used with the CASPER tools \citep{2016JAI.....541001H}), each processing a subset of frequencies for all antennas. Samples are aligned by a time-stamp before addition. A per-antenna real gain is provided for beam shaping; this is generally left at unity, but can be set to zero to eliminate an antenna from the beam, perhaps if it is not working properly. The B-engine output is also an 8-bit complex number and uses a requantisation gain to scale down the sum of the antenna voltages; this is typically scaled by $1/\sqrt{N_{\rm ants}}$. The output data is sent back onto the switch as {\sc spead}\footnote{https://casper.ssl.berkeley.edu/wiki/SPEAD} streams, one for each polarisation, using UDP multicast for consumption by downstream users. The packet sizes and rates are listed in Table  \ref{tb:bengine-configuration}. The B-engine can produce up to four tied-array beams from up to four simultaneous sub-arrays for downstream processing by the PTUSE pulsar processing servers. The computational burden on each server is identical regardless of the number of antennas in the sub-array.

\begin{table}
    \caption{MeerKAT B-Engine output configurations, based on the F-engine sampling interval and data rates.}
    \label{tb:bengine-configuration}
    \begin{tabular}{llll}
        \hline
        Band & Channels & Packet Size & Packet Rate \\
         & & (B) & (kPackets/s) \\
        \hline
        L & 1024 & 2048 & 1671.8 \\
        L & 4096 & 4096 & 835.9 \\
        UHF & 1024 & 2048 & 1062.5 \\
        UHF & 4096 & 4096 & 531.3 \\
        \hline \\
    \end{tabular}
\end{table}

\subsection{PTUSE: An SKA Pulsar Processing Prototype}
\label{sec:PTUSE}
PTUSE stands for Pulsar Timing User Supplied Equipment in the standard SARAO nomenclature. This sub-system receives channelised voltage timeseries from the B-engines. Each of the tied-array beams are received on separate high end server class machines and processed to produce reduced data products which are then transferred to the MeerKAT data archive for long term storage and subsequent processing. The system design was developed by Swinburne University of Technology as the pulsar timing prototype for Square Kilometre Array (SKA) pre-construction.

Two commissioning servers were deployed in 2015 for development and early science activities and were used until December 2019. Four production servers were then deployed to be used for the MeerTime key science program, allowing for increased processing capabilities and simultaneous processing of 4 tied-array beams. The configuration of servers for each deployment is described in Table \ref{tb:ptuse-hardware-deployments}.

\begin{table*}
    \caption{PTUSE Hardware Deployments, detailing the hardware configuration of the commissioning and deployment systems used with MeerTIME.}
    \label{tb:ptuse-hardware-deployments}
    \begin{tabular}{lll}
        \hline
        Hardware & Commissioning System & Production System \\
        \hline
        CPUs & 2 x Intel E5-2623 v3 & 2 x Intel Silver 4110 \\
        RAM & 128 GB DDR4-2133 MHz & 192 GB DDR4-2666 MHz \\
        GPUs & 2 x NVidia Titan X (Maxwell) & 2 x NVidia 2080Ti \\
        NVME Disk System & N/A & 8 TB RAID (4 x 2TB Intel P4510) \\
        SATA Disk System & 12 TB RAID (4 x 4TB SATA) & 24 TB RAID (4 x 8TB SATA) \\
        40Gb NIC & Mellanox ConnectX3 & Mellanox ConnectX5 \\
        \hline \\
    \end{tabular}
\end{table*}

Each PTUSE server subscribes to two {\sc spead} streams from the B-engines via a tree of Mellanox 40\,Gb switches, one for each polarisation. The {\sc spip}\footnote{https://github.com/ajameson/spip} software library receives the two streams, merging them and writing them to a {\sc psrdada}\footnote{https://psrdada.sourceforge.net} ring buffer in CPU memory. The ring buffer is configured to hold approximately 20 seconds of data, providing buffer space to absorb any downstream processing lag. The ring buffer uses shared memory to facilitate asynchronous I/O between the smaller, faster writes of the UDP receive process and the larger, slower reads of the signal processing software. Monitoring software can also periodically sample the data streams to provide signal displays and diagnostics.

Regardless of the number of channels, the data are split into two equal sub-bands which are independently processed in parallel by the pipelines in the {\sc dspsr} \citep{vb11} software library. The pipelines perform the major signal processing functions on the GPUs which provide sufficient performance to process sub-bands in real-time. {\sc dspsr} performs coherent dedispersion and can produce folded pulsar profiles (fold mode) or filterbanks (filterbank mode), significantly reducing the output data rate. The fold mode and filterbank mode both support flexible configuration parameters defining the output data resolutions subject to the limits listed in Table \ref{tb:ptuse-processing-capabilities}.

\begin{table*}
    \caption{PTUSE Processing Capabilities}
    \label{tb:ptuse-processing-capabilities}
    \begin{threeparttable}
    \begin{tabular}{llll}
        \hline
        Parameter & Fold Mode & Filterbank Mode & Unit \\
        \hline
        pulsar ephemeris & catalogue or custom & N/A & \\
        dispersion measure & 0 to 2000 & 0 to 2000 & pc cm$^{-3}$ \\
        output phase bins & 64 to 4096 & N/A &  \\
        output polarisation products $\dagger$ & 1, 2 or 4 & 1, 2 or 4 \\
        output sampling interval & N/A & (8 to 1024) * Tsamp & microseconds \\
        output sub-integration length & 8 to 60 & N/A & seconds \\
        output quantisation & 16 & 1, 2, 4 or 8 & bits per sample \\
        \hline
    \end{tabular}
        \begin{tablenotes} 
    \item[$\dagger$] With polarisations H and V, 1 denotes Stokes I (HH + VV), 2 denotes the square law detected power for each polarisation (HH, VV), 4 denotes the square law detected power in each polarisation and the real and imaginary components of the covariance between the polarisations (HH, VV, Real(H*V), Imag(H*V)).
    \end{tablenotes}
    \end{threeparttable}
\end{table*}

Both fold-mode and filterbank-mode data write the sub-banded results to disk in {\sc psrfits} \citep{hvm04} format, which are subsequently combined into a single file and transferred to both the MeerKAT Data Archive and the Swinburne OzSTAR supercomputer. The volume of filterbank data can be extreme when observing at the highest filterbank time resolutions, typical of globular cluster observing, recording data at over 400\,MB/s (30\,TB/day). These data products may be reduced on machines on-site prior to transfer to the data archive but also copied to future mirror sites planned in Europe (e.g. Max Planck Institute f\"ur RadioAstronomie).

\subsection{PTUSE: Challenges and Upgrades}
\label{sec:PTUSEsubsec}

During commissioning, it was found that capture of small UDP multicast packets at the high rates required is challenging on the Intel E5-2623v3 CPUs in the Commissioning System. The Commissioning System suffered from occasional packet loss, with an average loss rate of 700 bytes per minute (less than about 4 parts per billion) when observing at L-band in 1024 channel mode. Rather than having "holes" in the data, the samples from the previous cycle of the ring buffer were used to maintain the system noise level. This can occasionally replace what should be system noise by a pulse and vice versa. The issue was traced to insufficient CPU memory bandwidth, and to eliminate potential artifacts, the switch to the Production System has eliminated virtually all packet loss.

PTUSE supports a raw baseband observing mode which records the raw channelised voltage time series produced by the B-engines. The Commissioning Systems could only store $<$30\,s of data in this mode which would then require many minutes to slowly write out to the SATA disk system. The Production Systems each feature an NVME RAID disk which can record at 6 GB/s, which adds the capability to record up to 40 minutes of raw baseband data to disk on each server. This baseband mode allows recording at the native 1.196 $\mu$s time resolution of the PFB channel bandwidths, enabling the study of giant pulses, pulse microstructure or the timing of many globular cluster pulsars at Nyquist time resolution upon playback. 

The Commissioning Systems will be retained for data distribution and monitoring with the Production Systems used for all future PTUSE observations.

The {\sc dspsr} software library supports the creation of multiple frequency channels within each coarse channel (using the -F option on the command line) but to date there has been no call for it from observers and the user interface
does not yet support it. Writing baseband data to disk and running {\sc dspsr} on the command line would achieve this functionality if required.

\section{System Verification}
\label{sec:SV}

\subsection{System Equivalent Flux Density}
\label{SEFD}
Most pulsars at high dispersion measure (i.e. with DM > 200\,pc\,cm$^{-3}$) show only modest ($<$ 1 dB) flux density variation at the MeerKAT observing frequencies and can be used to perform a first-order calibration of the system performance. Three such pulsars are PSRs~J1602$-$5100 (B1558$-$50), J1651$-$4246 (B1648$-$42) and J1809$-$1917 with flux densities of 7.0, 21.4 and 2.8 mJy at 1369~MHz as measured by the Parkes telescope \citep{2018MNRAS.474.4629J}. 
Each of these pulsars was observed on 4 separate occasions with MeerKAT using the L-band receiver and here we investigate the system performance they imply for the telescope. The data were excised of interference and split into two bands each of 194~MHz centered near 1200 and 1400~MHz in order to best compare with the Parkes data. We derived a system equivalent flux density (including the sky contribution) of 8.1, 8.0, and 10.3~Jy in the central part of the band for the three pulsar locations. It is difficult to accurately estimate the sky contribution with the current MeerKAT system, but the Parkes observations would imply values of 4, 4, and 10~K in the direction of the three pulsars. We therefore conclude that the \textrm{SEFD} is consistent with 7~Jy across 400~MHz of bandwidth.

\subsection {Bandwidth Utilisation}

\begin{figure}
    \centering
    \includegraphics[width=\columnwidth]{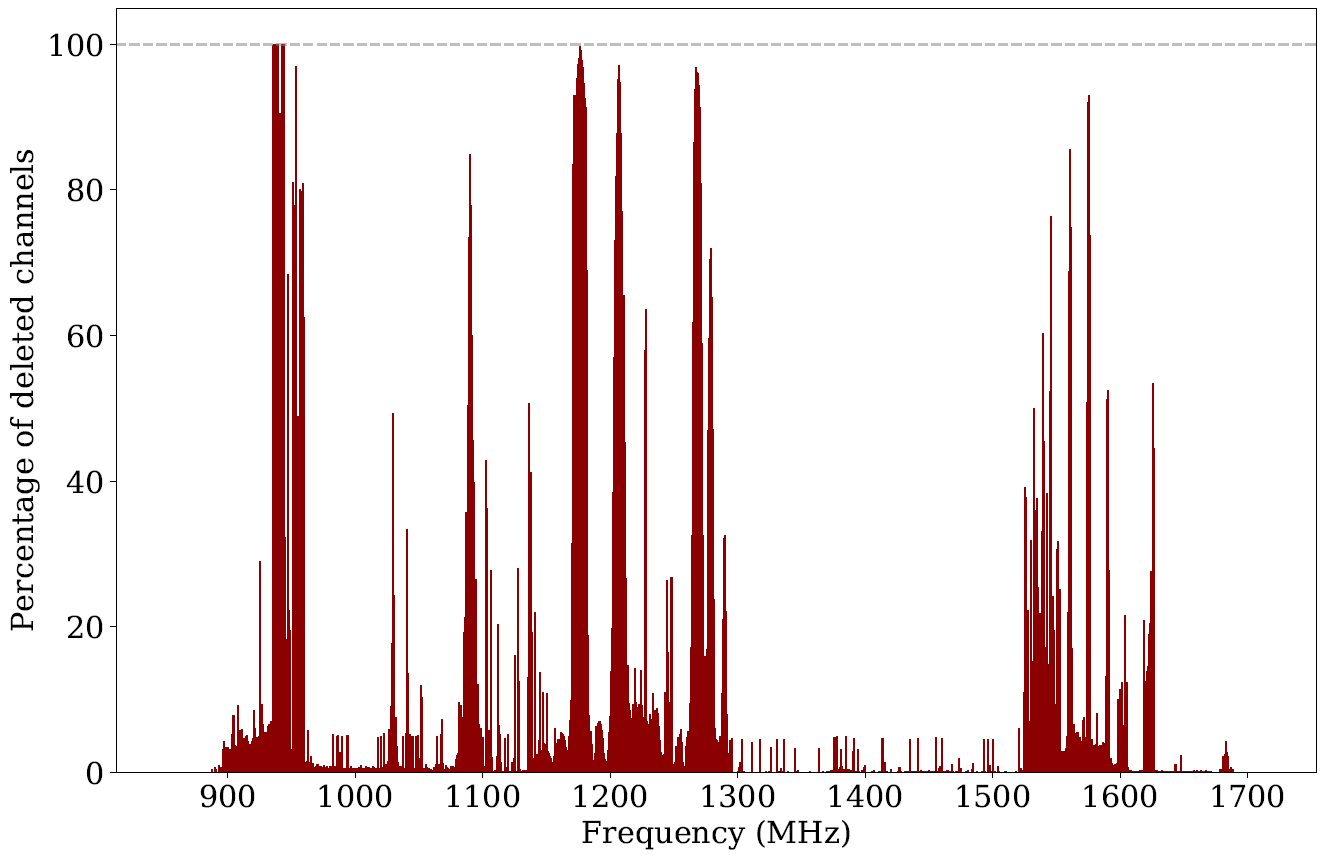}
    \caption{The fraction of 8-second folded integrations on PSR~J1909$-$3744 
    where the baseline
    had an integrated boxcar greater than 5\,$\sigma$ from the mean and were
    consequently deleted between Feb-Nov 2019 using the L-Band receiver.}
    \label{fig:RFIplot}
\end{figure}

To quantify the effect of radio frequency interference (RFI) on our pulsar science,
we took observations of the narrow duty-cycle MSP
PSR~J1909$-$3744 from Feb 2019 until Nov 2019 and eliminated 
interference by looking for deviations in the pulsar's baseline
that exceeded 5 sigma after averaging the pulse profile every
8 seconds. In total, we analysed 2825 8-second integrations.
These results are shown in Figure~\ref{fig:RFIplot}.
In the central 928 channels of the 1024, we only had to delete 
12.8\% of the 1400 MHz band, on average. Although, some of these channels have very persistent RFI and their deleted fraction is close to 100\%, particularly those
channels associated with Global Positioning Satellites and
the mobile phone band near 950 MHz. Other RFI sources
are very time-dependent like those associated with aircraft.

On short timescales, the fraction of affected integrations is similar.
In an analysis of a 7200-second observation of the giant-pulse
emitting pulsar PSR~J0540$-$6919 \citep[B0540$-$69][]{2003ApJ...590L..95J}, we created single pulse
timing archives with an approximate integration time of 50.6\,ms.
We followed the same procedure we used for the PSR~J1909$-$3744 observations to delete RFI-affected frequency channels.
Single-pulse integrations make weaker RFI easier to detect as it is not washed out by the process of pulsar folding but also means RFI with an on/off timescale greater than (in this case) 50\,ms can lead to integrations with less or almost no RFI. We found that, in a single two-hour observation, 9.6\% of the band was deleted using the same criteria as for the integrated pulse profile tests.

To place these results in context, for much of the last two decades, observations at the Parkes 64\,m telescope have used bandwidths of 256-340\,MHz in the 20-cm band. Our results suggest that, for pulsar timing and single-pulse
studies, effectively 87-90\% of the 928 central frequency channels 
can be used for a total bandwidth of 675-700\,MHz centred at 1400\,MHz.
This is very competitive with almost all existing large-aperture telescopes
and comparable to the fraction of the 1400\,MHz band at the NRAO Green Bank telescope (GBT) and only exceeded by the recent development of the Ultra-Wideband receiver \citep{2019arXiv191100656H} at the Parkes telescope that operates from 704\,MHz to 4.032\,GHz.
In Figure 6 we show the examples of broadband observations of the double
pulsar PSR J0737--3039A that demonstrate the high fractional bandwidths
available for pulsar observations at UHF and L-band with MeerKAT.

\begin{figure*}
    \centering
    \begin{subfigure}{0.44\textwidth}
        \includegraphics[angle=0,width=\columnwidth]{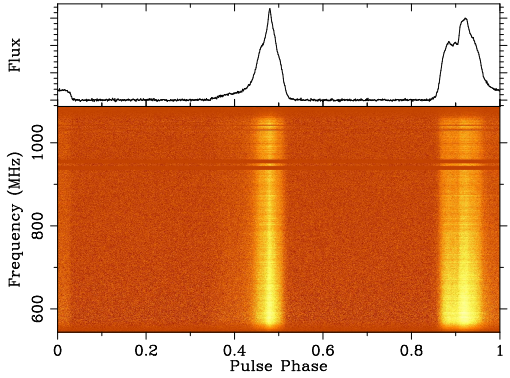}
        \caption{Integrated pulse profile for PSR~J0737$-$3039A
        with the UHF receiver. The flux density scale is
        arbitrary.}
       \label{fig:UHF_profile}
    \end{subfigure}
    \hfill
    \begin{subfigure}{0.44\textwidth}
        \includegraphics[angle=0,width=\columnwidth]{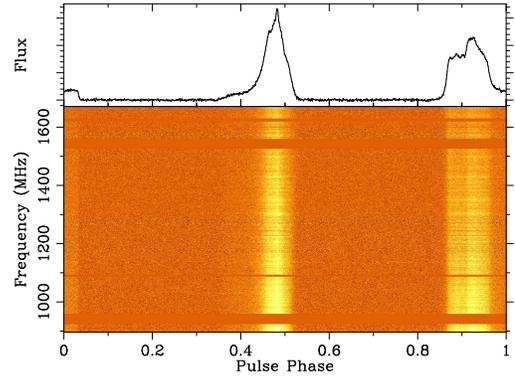}
        \caption{Integrated pulse profile for PSR~J0737$-$3039A with the L-band receiver. The flux density scale is arbitrary.}
        \label{fig:Lband_profile}
    \end{subfigure}
    \caption{
    Observations of PSR~J0737$-$3039A with MeerKAT's UHF and L-Band receivers.
    }
    \label{fig:j0737obs}
\end{figure*}

\subsection{Spectral Leakage}
\label{leakage}

\begin{figure}
    \centering
    \includegraphics[width=\columnwidth]{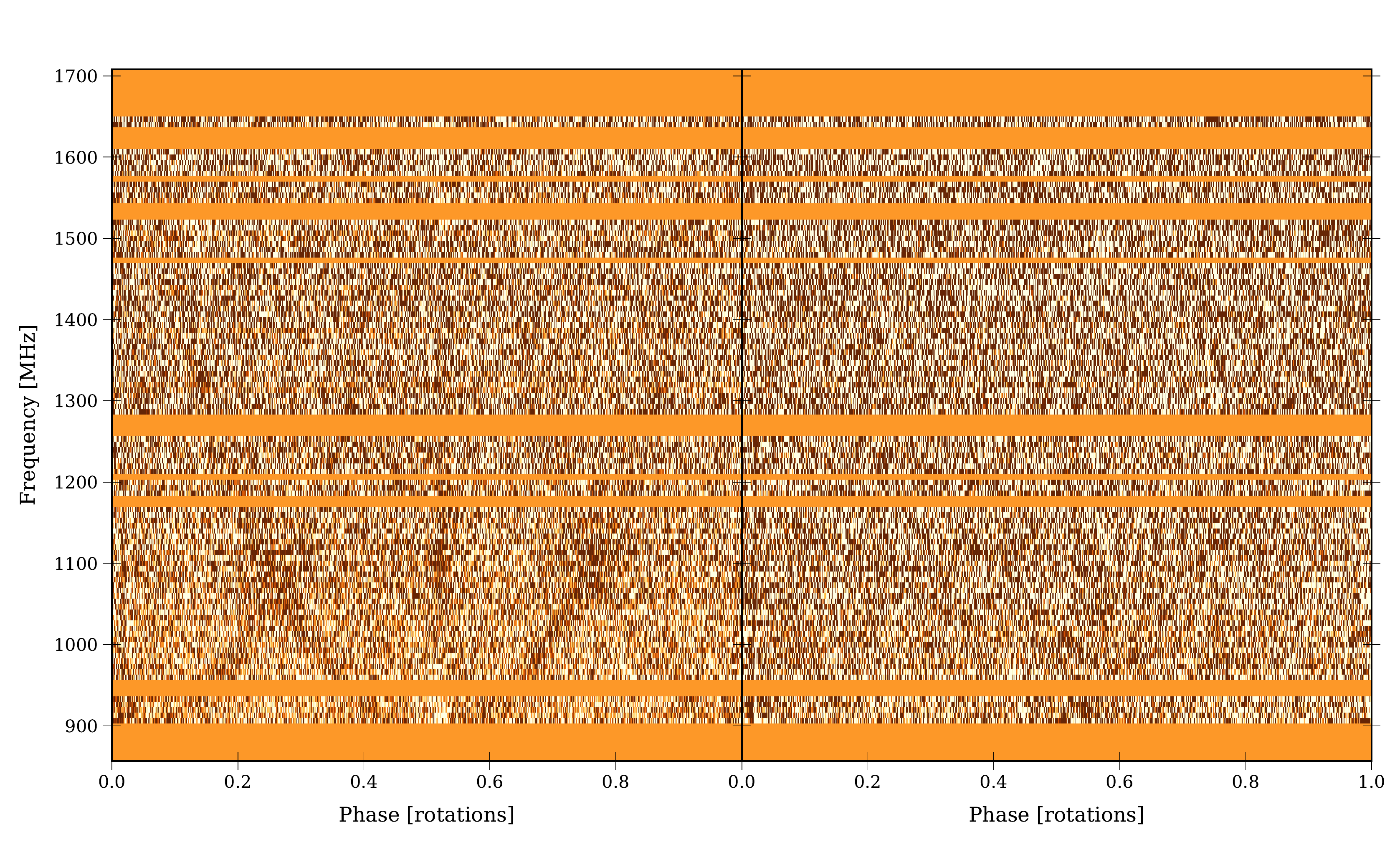}
    \caption{Average pulsar profiles after the bright main pulses and weaker interpulse are subtracted using a frequency-dependent mean analytical profile. This reveals the extent of the artifacts that were present in
    the original filters (left panel) and the extent to which they have been removed with the 0.91 filter design (right panel).}
    \label{fig:pennuccimask}
\end{figure}

The effect of spectral leakage, due to the shape of the PFB filter magnitude response in the F-engines, is demonstrated with 700-second observations of PSR~J1939+2134 (B1937+21). These observations were coherently dedispersed, folded and integrated into a single profile.  The difference between an average frequency-dependent profile for the two filter designs (described in section \ref{PFB} ) as a function of frequency and pulse phase is shown in Figure \ref{fig:pennuccimask}. The original filters gave rise to frequency-dependent pulse profiles that possessed `reflections' of the main and inter-pulses due to spectral leakage from the adjacent channels. At lower frequencies the reflected pulses are offset further from the true pulse with the magnitude of the reflection being inversely proportional to the amplitude response at the channel boundary in the filter design. This is very bad for precision timing experiments, as depending upon the location of scintillation maxima it can systematically alter the shape of the pulsar's frequency-integrated profile and lead to  systematic timing errors. The new filter greatly reduces the amplitude of these artifacts which are now seemingly negligible.

\subsection{Artifacts}
\label{ART}
To explore the level of any potential system artifacts we compared the pulse profile of the MSP PSR~J1939+2134 observed with MeerKAT/PTUSE with archival 
observations from the CASPSR (CASPER-Parkes-Swinburne-Recorder) coherent dedisperser on the Parkes 64\,m radio telescope in the same frequency band. CASPSR digitizes the entire down-converted 400 MHz band and uses the \textsc{dspsr} library 
to coherently dedisperse the data using graphics processing units.
PTUSE on the other hand dedisperses the narrow polyphase filterbank channels produced by the B-engine using the same software library. 
There is perhaps a danger that each of these methods may create 
different artifacts in the profile that affect precision timing and interpretation of pulse features.

\begin{figure}
    \centering
    \includegraphics[width=\columnwidth]{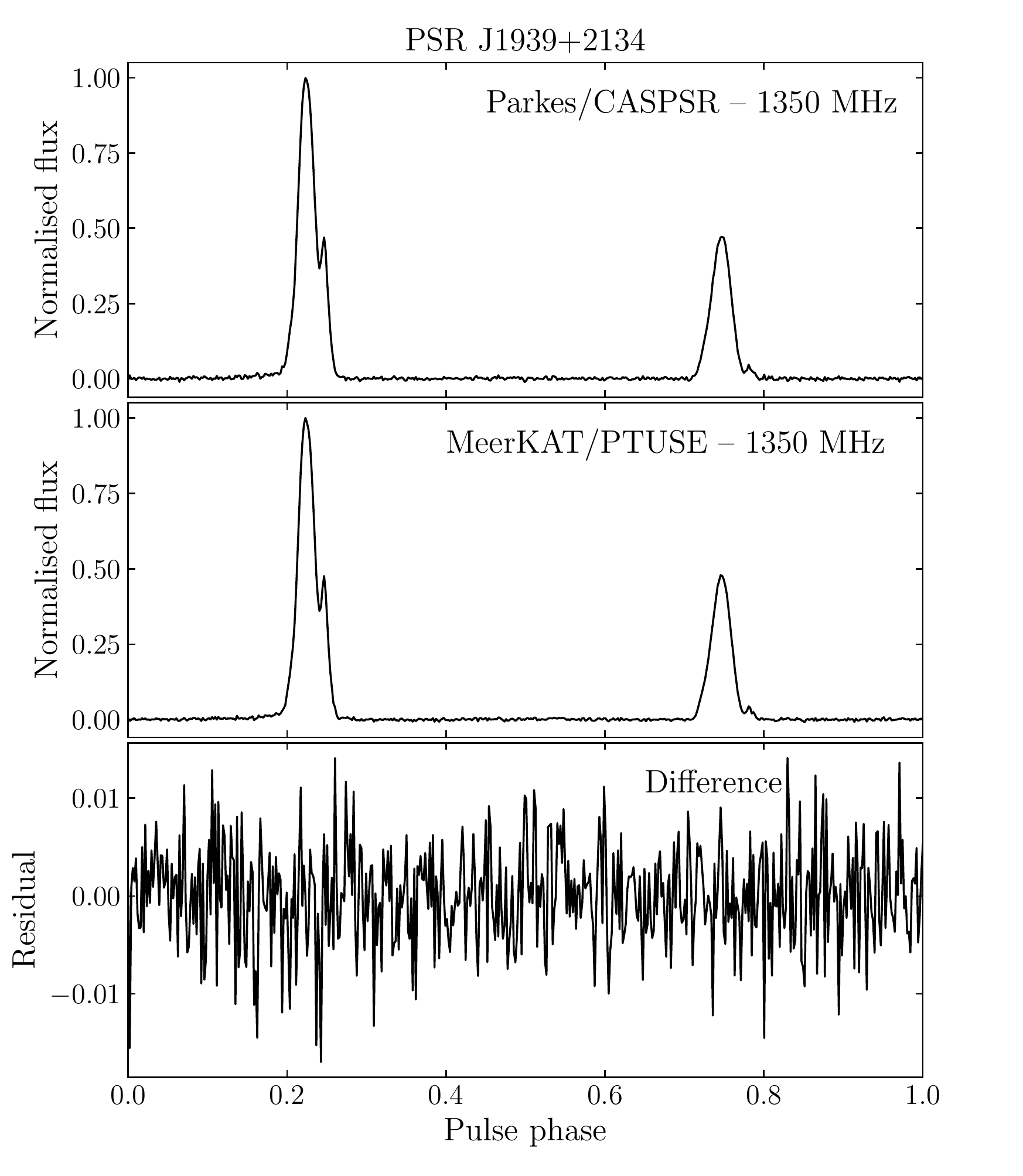}
    \caption{Observations in the same frequency band of PSR J1939+2134 at Parkes (top panel),
    MeerKAT (middle panel) and their difference (bottom panel) in normalised units to the profile peak. The agreement between the telescopes and backends is excellent.
    }
    \label{fig:1939_PKS_MK}
\end{figure}

PSR~J1939+2134 has a steep spectrum and is prone to strong scintillation maxima in narrow (few MHz) frequency bands around 1400\,MHz hence it is important to focus on a relatively narrow fractional bandwidth to identify
potential artifacts. We selected the relatively narrow bandwidth between 1280 and 1420 MHz at both sites and produced a Stokes I profile for each. As we saw earlier (section \ref{PFB}) in some MeerKAT F-engine modes (eg the early 1024 channel mode)
the spectral leakage is significant between neighbouring channels and the relative heights of the two pulse components from the MeerKAT profile did not agree with the CASPSR data with the weaker of the two pulses being reduced in amplitude by circa 10\%. We then repeated the exercise using the new 1K mode of MeerKAT/PTUSE with the sharper filters and found the MeerKAT and Parkes profiles to be consistent
within the noise. 
We attribute this improvement to the choice of filters now in use in MeerKAT's F-engine that eliminate spectral leakage.
The Parkes, MeerKAT and difference profiles are shown in Figure \ref{fig:1939_PKS_MK}.
This is encouraging for comparisons of pulsar profiles not only between different pulsars at MeerKAT but between observatories that use digital signal processing and common 
software libraries.

\subsection{Polarimetry}
\label{POLN}
The boresight polarimetric response of the MeerKAT tied-array beam was estimated using the Measurement Equation Modeling (MEM) technique described in \citet{van04}.  Motivated by the results of \citet{lck+16}, the MEM implementation was updated to optionally include observations of an artificial noise source that is coupled after the orthomode transducer (OMT) and to remove the assumption that the system noise has insignificant circular polarisation. The updated model was fit to observations of the closest and brightest MSP, PSR~J0437$-$4715, made over a wide range of parallactic angles, and both on-source and off-source observations of the bright calibrator PKS~J1934$-$6342.

The best-fit model parameters include estimated receptor ellipticities that are less than $1^\circ$ across the entire band, indicating that the degree of mixing between linear and circular polarisation is exceptionally low. The non-orthogonality of the receptors is also very low, as 
characterised by the intrinsic cross-polarisation ratio \cite[IXR;][]{cw11}, which varies between 50 and 80\,dB across the band. Noting that larger values of IXR correspond to greater polarimetric purity, the MeerKAT tied-array beam exceeds both the minimum pre-calibration performance \citep[$\sim 30$\,dB;][]{fkp+15} and the minimum post-calibration performance \citep[$\sim 40$\,dB;][]{ckl+04,van13} recommended for high-precision pulsar timing.

The reference signal produced by the incoherent sum of the noise diode signals from each antenna significantly deviates from 100\% linear polarisation; its polarisation state varies approximately linearly from $\sim20$\% circular polarisation at 900~MHz to $\sim60$\% circular polarisation at 1670~MHz. Therefore, if an observation of the reference signal were to be used to calibrate the differential gain and phase of the tied-array response, then the technique described in Section 2.1 of \cite{ovhb04} would be necessary.  

However, the reference signal also exhibits evidence of a significantly non-linear tied-array response. This is observed as over-polarisation of the reference signal (e.g., degree of polarisation as high as 105\% -- 110\%) and is also observed in the goodness-of-fit (e.g.\ reduced $\chi^2$ between $\sim$300 and $\sim$800) reported when performing MEM with reference source observations included.  The origin of the non-linearity is currently not understood; therefore, given that the best-fit values of differential receptor ellipticity are very small\footnote{Differential receptor ellipticity, which describes the mixing between Stokes~I and Stokes~V, must be constrained by observations of a source of known circular polarisation, as described in Appendix B of \citet{van04} and considered in more detail in \citet{lck+16}.}, all reference source observations (including on-source and off-source observations of PKS~J1934$-$6342) were removed from the MEM input data, yielding good fits to the pulsar signal with reduced $\chi^2$ between $\sim$1.6 and $\sim$1.9.

\begin{figure*}
    \centering
    \includegraphics[angle=-90,width=0.75\textwidth]{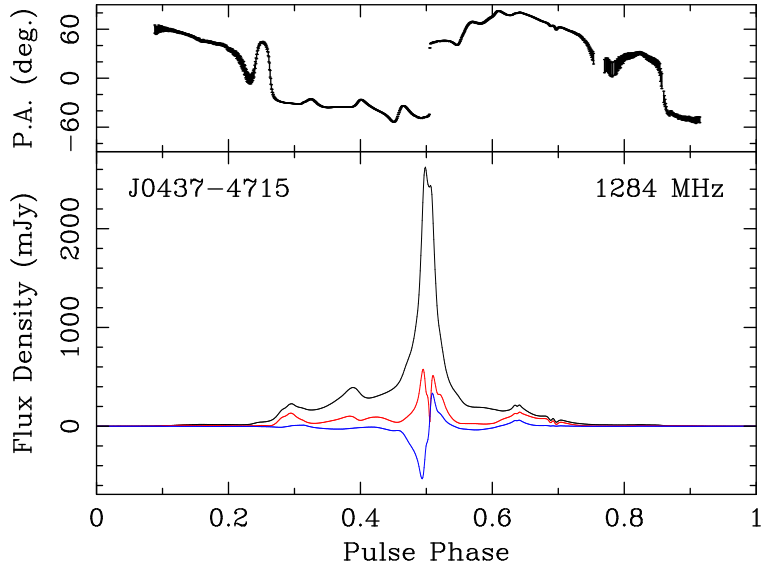}
    \caption{Calibrated polarisation of PSR~J0437$-$4715, plotted as a function of pulse phase.
    In the top panel, the position angle of the linearly polarized flux is plotted with error bars indicating 1 standard deviation.  In the bottom panel, the total intensity, linear polarisation, and circular polarisation are plotted in black, red and blue, respectively.}
    \label{fig:0437_poln}
\end{figure*}

To test the stability of the polarimetric response, observations of PSR~J0437$-$4715 made on 4 September 2019 were modelled and calibrated using MEM and then integrated to form a template with which to model observations made on 3 October 2019 using Measurement Equation Template Matching \cite[METM;][]{van13}.  
%
%
The template, formed from an integrated total of 2 hours of observing time, has a signal-to-noise ratio of $3.8 \times 10^4$.  In each frequency channel that was not flagged as corrupted by RFI, the METM model fit the data well, with reduced $\chi^2$ values ranging between $\sim$1.1 and $\sim$1.3.  The integrated total of the METM-calibrated data are plotted in Figure~\ref{fig:0437_poln}.

As a final consistency check, the calibrated polarisation of PSR~J0437$-$4715 observed at MeerKAT was quantitatively compared with that observed at the Parkes Observatory using   CASPSR.  After selecting the part of the MeerKAT band that overlaps with the 400~MHz band recorded by CASPSR, the calibrated MeerKAT data were fit to the calibrated Parkes template using Matrix Template Matching \cite[MTM;][]{van06}.  The MeerKAT data fit the Parkes data well, with reduced $\chi^2$ values ranging between $\sim$1.2 and $\sim$1.5.  The Jones matrices that transform the MeerKAT data to the basis defined by the Parkes template in each frequency channel were parameterised using Equation 19 of \cite{bri00}.
All model parameters were close to zero, except for the differential ellipticity, $\delta_\chi$, which varied between +1 and $-$2\,degrees as a function of frequency, and the rotation about the line of sight, $\sigma_\theta$, which varied between $-$5 and $-$8\,degrees.  Non-zero values of $\delta_\chi$, which describes the mixing between Stokes~I and Stokes~V, are expected; as described in Appendix B of \citet{van04}, this mixing must be constrained by introducing assumptions that may be correct only to first order.  
Non-zero values of $\sigma_\theta$ are also expected owing to unmodelled Faraday rotation in Earth's ionosphere.

After initialising the array, a standard operating procedure is to run the so-called delay calibration observation. During the observation, the noise diode as well as bright, well known sources are used to calculate and apply time-variable solutions for the antenna-based delays.
The delay calibration observation consists of multiple stages: initially predefined F-engine complex gains are applied in the correlator for each antenna; a suitable calibrator is observed and simple antenna-based delays are calculated; next, a noise diode is activated and cross-polarisation delay as well as phase is measured for the entire array. The delays are derived and combined by the real-time calibration pipeline before being applied to the data with the exception of the cross-polarisation phases which are stored in the observation metadata and can be applied at a later stage. At present the parallactic angle
is assumed to be the same for every antenna. Pulsars that transit 6 degrees from the zenith
have at most about a 0.2 degree error because of this assumption.


\subsection{Timing}
\label{TIMING}
To test the timing stability of the telescope, we routinely observed the bright, narrow MSP PSR~J1909$-$3744 over a period of about 11 months from
March 2019.

This pulsar has a well-established ephemeris that we took from the Parkes Pulsar Timing Array \citep[PPTA, ][]{kerr2020pptadr2}. 
We first excised radio frequency interference in the data cube using the {\sc coastguard} package \citep{lkg+16}, which we modified to work with MeerKAT data.  Importantly we used frequency-dependent model templates to 
identify on and off-pulse regions from which to calculate a set of 
statistics that could be used to identify contaminated profiles that 
should be excised.
We updated the dispersion measure in the data sets to a value near the mean over the observing interval.
We then averaged to $32$ channels in frequency and completely in time for each observation.  Using a template with $32$ frequency channels to capture profile evolution \citep[derived using the technique of ][]{Pennucci+19}, we derived arrival times using the Fourier-domain phase-gradient algorithm \citep{Taylor92} with Monte Carlo estimates for the arrival time uncertainties.  

We analysed the arrival times using {\sc temponest} \cite[][]{lah+14}. To first order, MSPs only drift slowly from their timing models, so we started the PPTA ephemeris from its second data release \citep{kerr2020pptadr2} and modelled  only the minimum number of additional parameters: the pulsar spin and spin down rate; dispersion measure and first derivative dispersion measure; and the {\sc tempo2} FD (``frequency dependent'') parameters to account for pulse profile evolution with frequency. FD parameters model systematic offsets in the average frequency-resolved timing residuals \citep{Arzoumanian2015b}.  The model is a simple polynomial as a function of log$_{10}$ of the observing frequency, where each FD parameter is one of the polynomial coefficients.  However, it should be noted that FD parameters can absorb other systematic effects besides unmodeled evolution of the profile shape.

We searched for three forms of stochastic noise in the data. 
We searched for red noise and DM variations using the established Bayesian methods now commonly employed. 
To model the white noise we searched for both EQUAD and EFAC (using the {\sc temponest} definitions), and a new parameter, TECORR, which accounts for correlated white noise, by adding an additional term to the noise covariance matrix
\begin{equation}
\sigma^2_{ij}=\delta(t_i-t_j) \sigma^2_{\rm TECORR, hr} \sqrt{3600~{\rm s}/T},   
\end{equation}
where $\delta(t_i-t_j)=1$ if the data are from the same integration, and $T$ is observation integration time in seconds.
The noise parameter is similar to the ECORR that has been employed notably in NANOGrav data analyses \citep[Alam et al. \textit{in prep.},][]{Arzoumanian18b,Arzoumanian2015b}.  However the noise accounts for varying observing lengths. Correlated noise introduced by stochasticity in pulse shape variations is predicted to reduce in proportion to the square root of time.
We find no evidence for red noise in our data set, which is unsurprising given its short length.
We note that the Bayesian methodology employed accounts for covariances between noise and the timing model, so would be able to detect red noise if it were sufficiently strong. 
We find strong evidence for dispersion measure variations, which are visible by eye in our $512$-s observations.
We also find evidence for band-correlated white noise (TECORR), but no evidence for EQUAD and EFAC.   
We measure $\sigma_{\rm TECORR,hr} \approx 24$\,ns. Which is approximately a factor of two larger than the expected jitter noise measurements inferred from our short observations  \cite[][]{sod+14}.  The excess noise is the subject of current research, but we suspect it includes contributions from unmodelled dispersion measure variations \cite[][]{css16,sc17}. 

\begin{figure}
    \centering
    \includegraphics[width=\columnwidth]{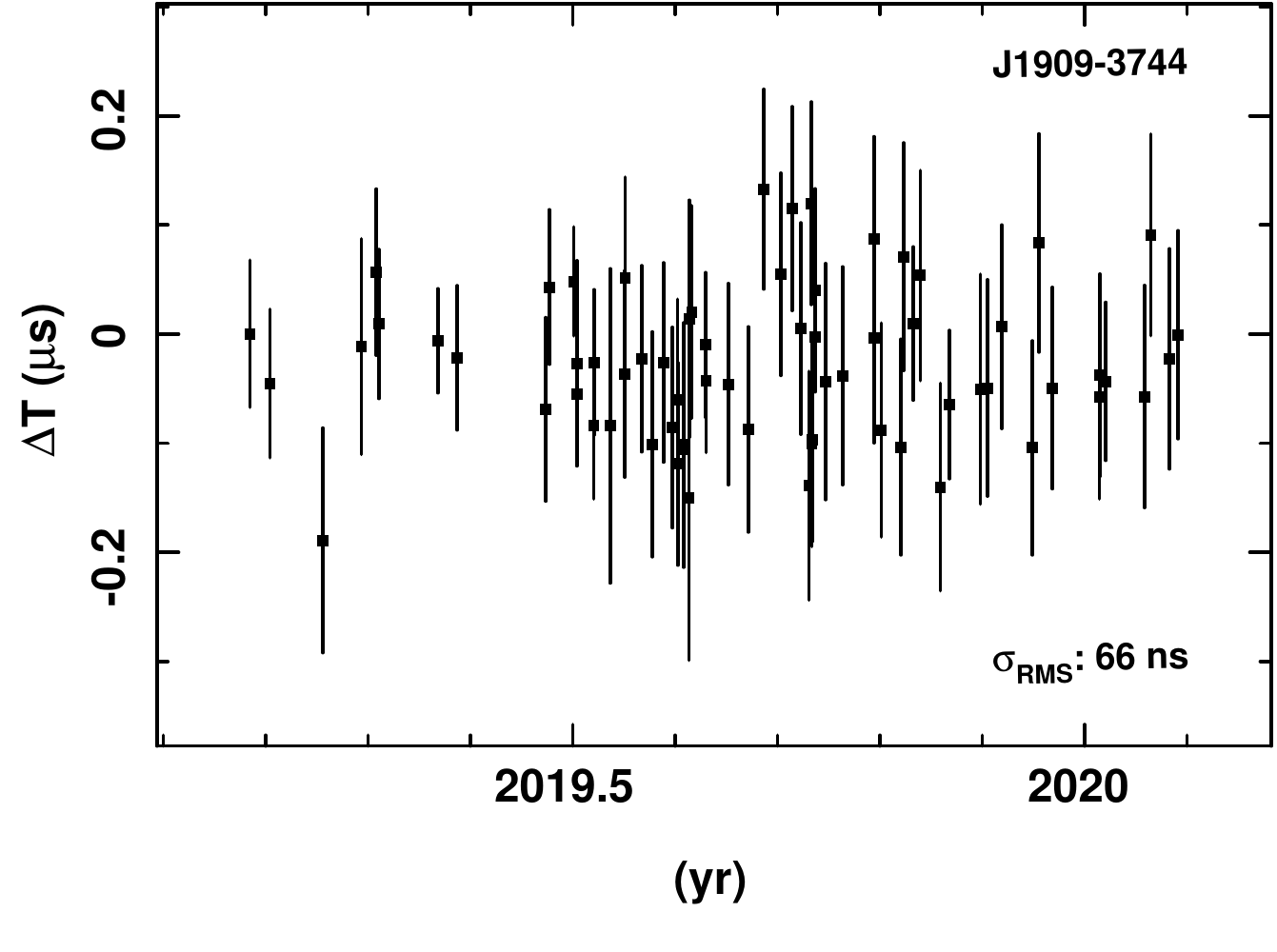}
    \caption{ Epoch-averaged residual arrival times for PSR~J1909$-$3744.    }
    \label{fig:1909timing}
\end{figure}

After subtracting the maximum likelihood model for dispersion measure variations and forming the weighted average of the sub-banded residual arrival times we found evidence for marginal orbital phase dependent variations in the residuals.  After accounting for this by fitting for the companion mass, we measure the root-mean square of the average residual arrival times times to be $\approx 66$\,ns as shown in Figure \ref{fig:1909timing}. It is possible that the fitting (in particular of the position and spin down) is absorbing some noise in the data set.
If we start from the ephemeris published in PPTA-DR2 \cite[][]{kerr2020pptadr2}, fitting only  DM variations and FD parameters we measure the rms of the averaged residuals to be 76\,ns. We note that the PPTA-DR2 ephemeris was intended to be initial ephemeris for future studies, and only crude noise modelling was undertaken. 

\begin{figure}
    \centering
    \includegraphics[width=\columnwidth]{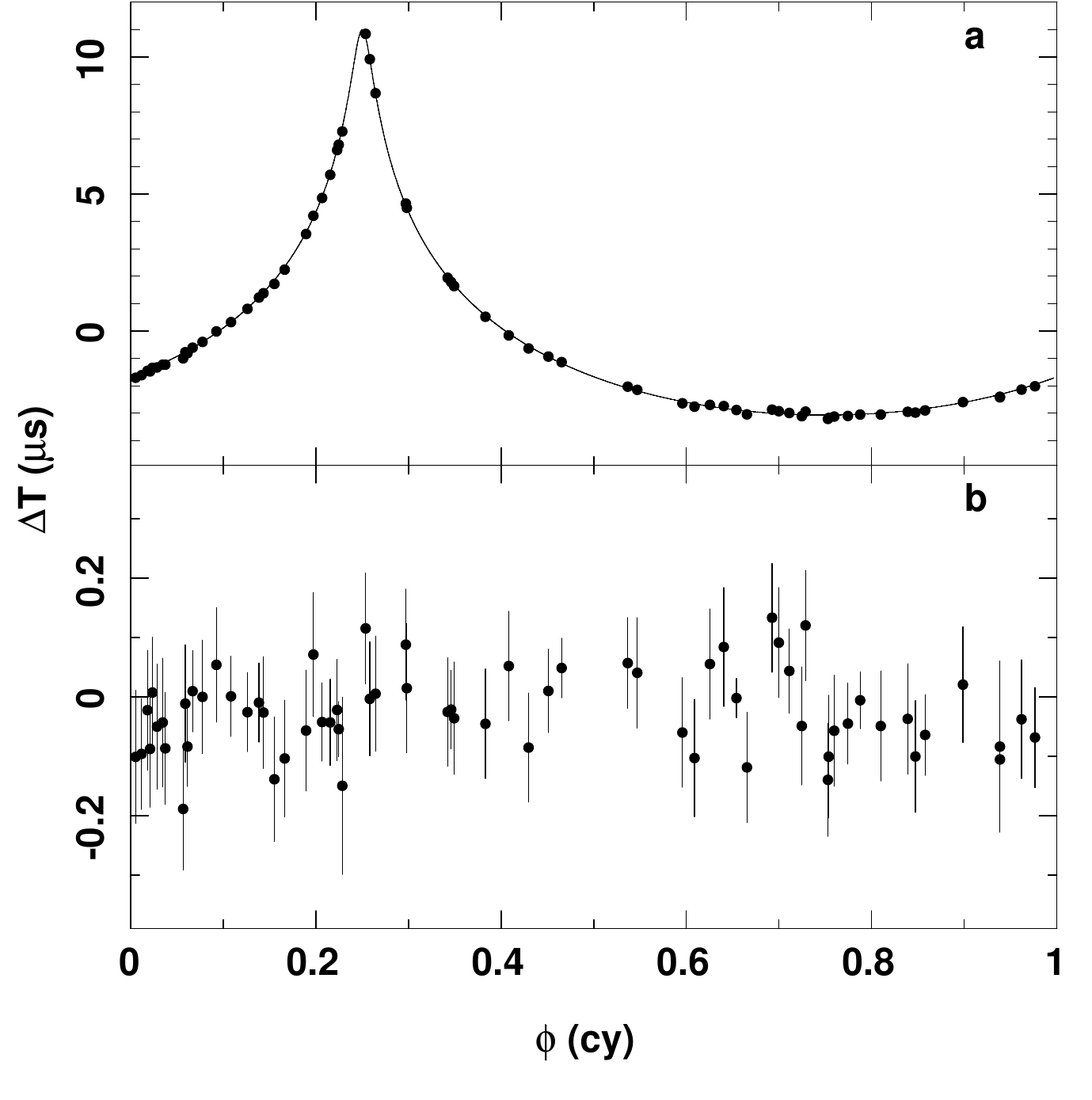}
    \caption{Epoch averaged residual arrival times for PSR~J1909$-$3744 plotted against orbital phase
    in cycles. Residual arrival times are plotted using the maximum-likelihood model without (panel a) and with accounting for the Shapiro delay induced by the companion. The predicted signal is shown as the solid line in panel a.}
    \label{fig:1909timing_orb}
\end{figure}

Figure \ref{fig:1909timing_orb} shows the residual arrival times plotted versus pulse phase when ignoring entirely (panel a) and accounting for the Shapiro delay caused by the radio waves propagating through the gravitational field of the companion.

To test the timing stability of the system on short timescales, we made use of the bright Fermi source PSR~J2241$-$5236 \citep{2011MNRAS.414.1292K}. This 2.18\,ms MSP has a narrow duty-cycle ($\sim$3\%) and is regularly timed as part of the Parkes Pulsar Timing Array. It has a mean 1.4\,GHz flux density of 3-4\,mJy and often experiences bright scintillation maxima. On April 22 2019 UTC,  this pulsar produced a 5030\,$\sigma$ profile in just 512\,s in 64 $\times$ 8\,s integrations.
After forming an appropriate smoothed template, we produced arrival times every 8\,s after summing over the full bandwidth and obtained an rms residual of the resultant 8\,s integrations of only 90\,ns using
the existing ephemeris. This is the lowest rms residual in 8\,s ever seen in pulsar timing and implies a very small jitter upper limit of only $90/(3600/8)^{1/2} = 4.2$\,ns in an hour. 
\begin{figure}
    \centering
    \includegraphics[width=\columnwidth]{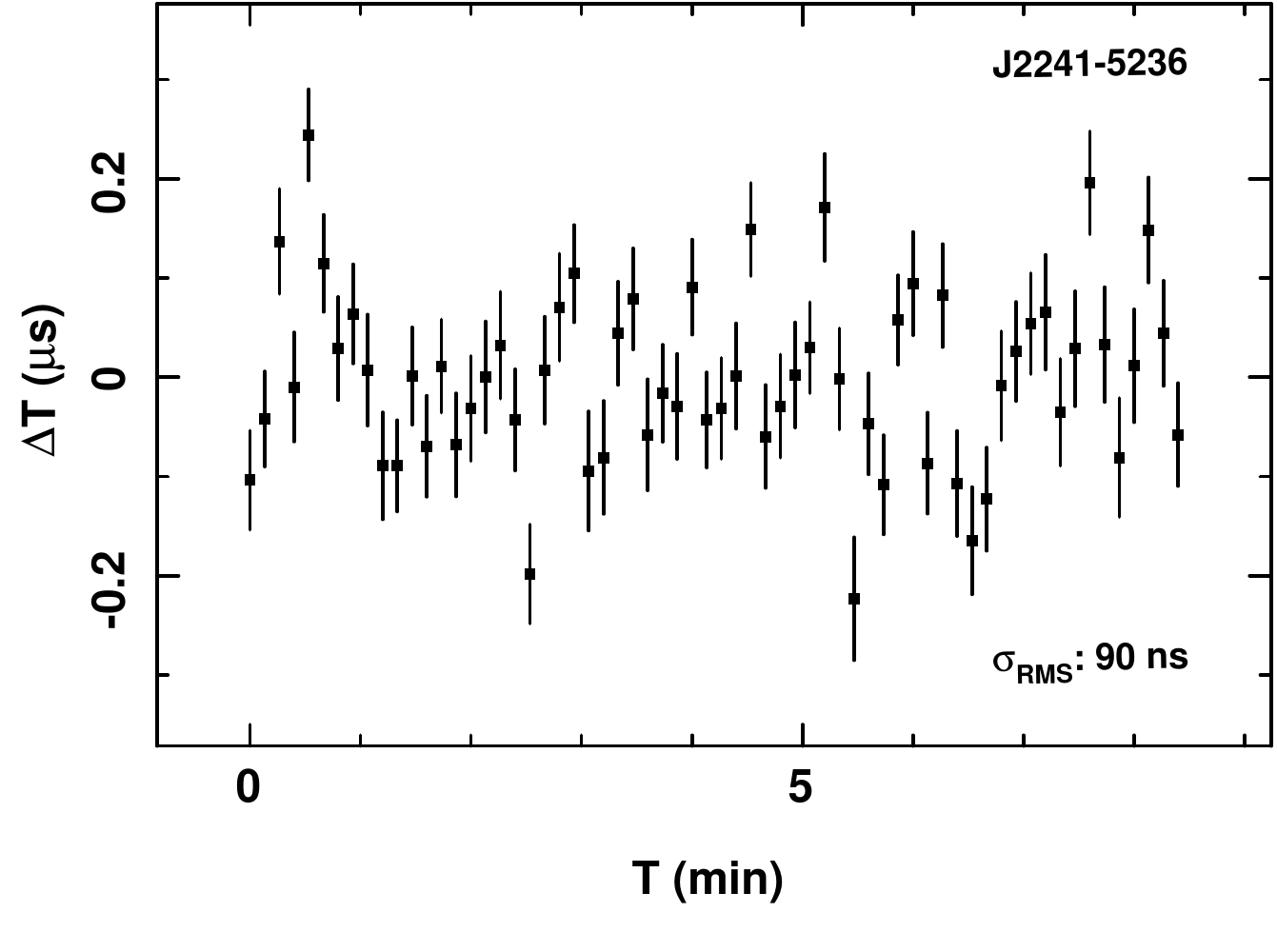}
    \caption{Times of arrival during a 512s observation of PSR~J2241$-$5236 
    with MeerKAT using the L-band receiver. The post-fit rms residual is
    just 90\,ns using 8-s integrations. This implies a jitter limit of less than
    4.2\,ns in one hour.}
    \label{fig:2241toas}
\end{figure}

Both of these results auger well for the future of MSP timing at the MeerKAT telescope and are a testament to the engineering care that has been achieved with the TFR, PTM and PTUSE.

\subsection{Filterbank Mode}
\label{subsec:GC}
\begin{figure}
    \centering
    \includegraphics[width=\columnwidth]{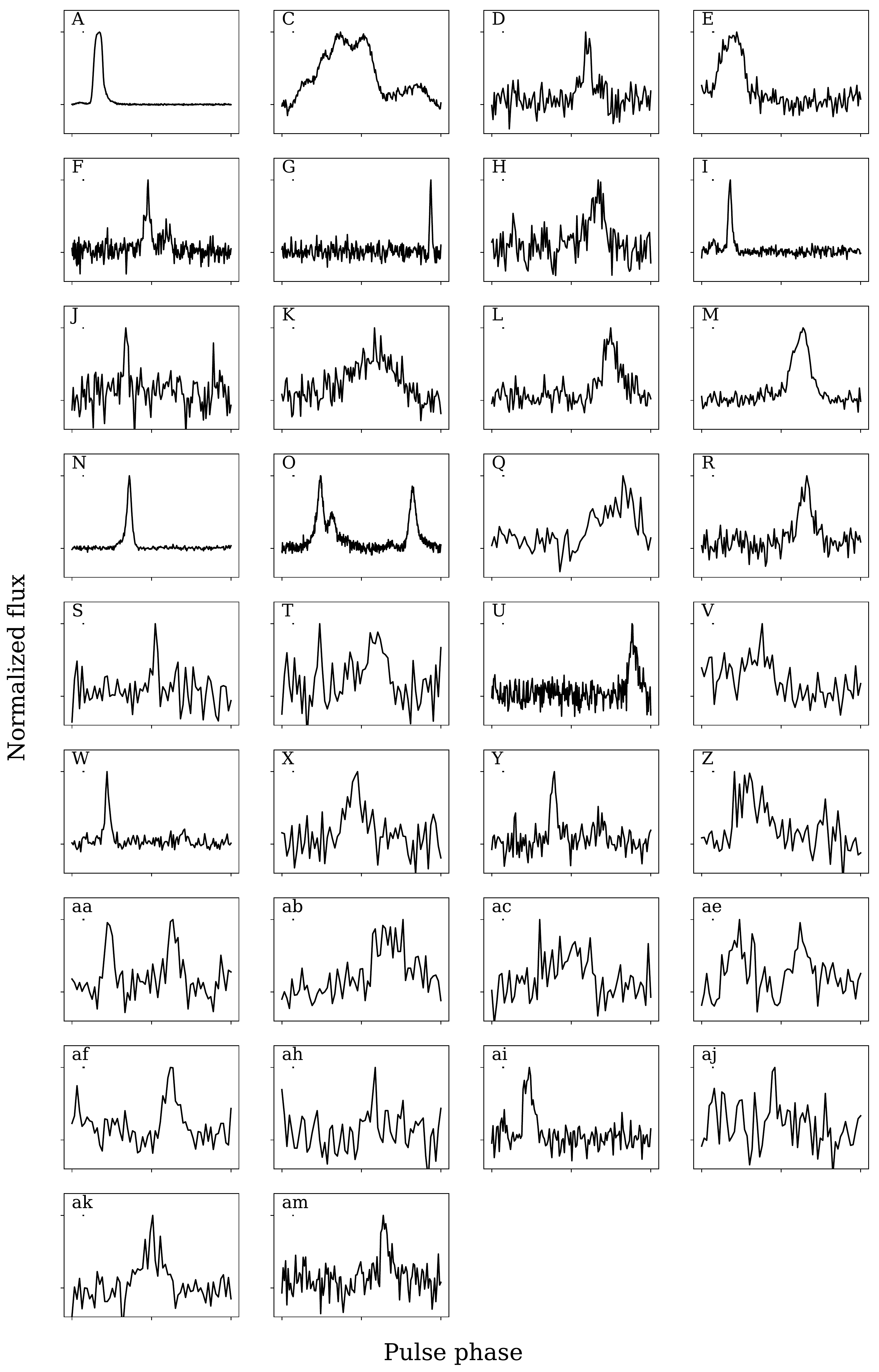}
    \caption{Folded pulse profiles of the 34 pulsars in Terzan\,5 from a 9000\,s integration.
    }
    \label{fig:Ter5montage}
\end{figure}

The globular cluster Ter\,5 was observed in filterbank mode for 9000\,s on May 27 2019 with 9.57\,$\mu$s time resolution and full polarimetry. Profiles for the 34 detected pulsars are shown in Figure \ref{fig:Ter5montage} showing the high time resolution 
and signal-to-noise ratios for many of the pulsars. These observations only used the central dishes within 500\,m radius of the core. A trade-off has to be made between
the sensitivity of the tied beam and its width. With this
configuration all of the pulsars 
except Ter\,5\,A, D, X and J were within the half-power
point of the beam at the centre frequency of the observation.
Ter\,5\,A is so bright that it was easily detected regardless, but the
others were heavily attenuated. 
The detections of Ter\,5\,ah and Ter\,5\,aj are marginal. 
The pulsar profiles were obtained in a two-stage process. First, the pulsars
were folded with the latest ephemerides available from the GBT program (Ransom, private communication). Some
required minor refinement of their periods to correct for minor drifts in pulse phase. It is not uncommon for
black widow pulsar systems to accumulate orbital phase drifts that manifest themselves as
changes in the observed period for fractions of their orbit. 
A comparison of S/N was made for 29 of the pulsars that
are routinely detected at the GBT in an equivalent observing time. 
Of these, 15 were better with MeerKAT and another 9
within 25\% of the GBT value. The four poorest (A, D, X and J) were all outside the half-power point of
the MeerKAT tied beam.

\begin{figure}
    \centering
    \includegraphics[angle=0,width=\columnwidth]{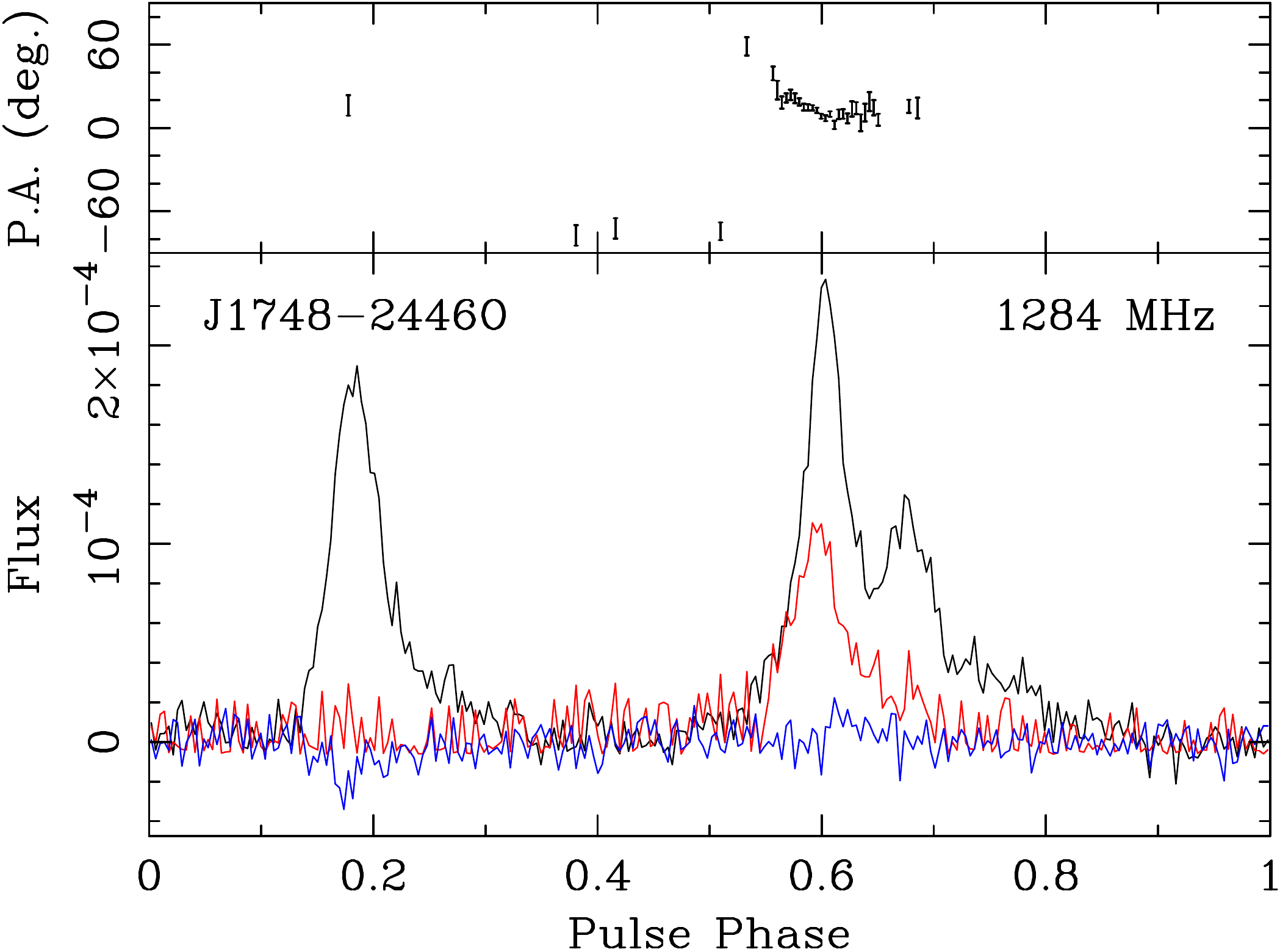}
    \caption{Calibrated polarisation profile of PSR~J1748$-$2446O, plotted as a function of pulse phase.
    In the top panel, the position angle of the linearly polarized flux is plotted with error bars indicating 1 standard deviation. 
    In the bottom panel, the total intensity, linear polarisation, and circular polarisation are plotted in black, red and blue, respectively.}
    \label{fig:Ter5O_poln}
\end{figure}

Ter\,5\,O was observed in parallel using the fold-mode of PTUSE and after calibration we derived a rotation measure for the cluster of 174.6$\pm$0.8 rad/m$^2$ which is consistent with the value (178.5$\pm$3.5) for the cluster obtained by \cite{2018ApJ...867...22Y} for Ter\,5\,A. The pulse profile and polarimetry are shown in Figure \ref{fig:Ter5O_poln}.

\section{New Results}
\label{sec:RESULTS}

\subsection{A Giant Pulse from PSR~J0540--6919}
\begin{figure}
    \centering
    \includegraphics[width=\columnwidth]{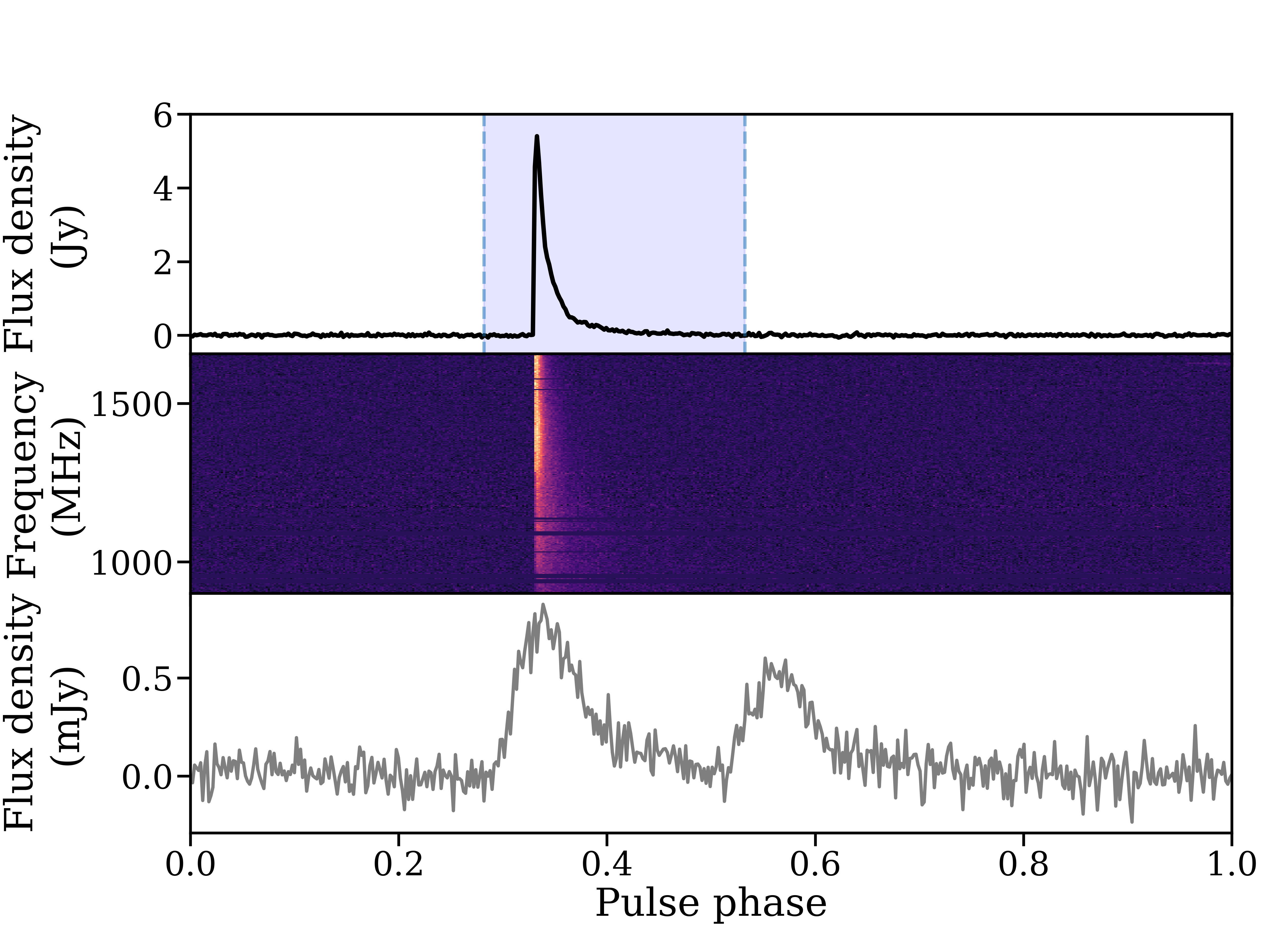}
    \caption{\textit{Top and middle:} Giant pulse from PSR~J0540$-$6919 with a peak flux density of 5.4\,Jy and an estimated mean flux density of $\sim$ 92\,~mJy. Using the SEFD of MeerKAT the off-pulse rms is estimated to be 20 mJy. The shaded region in the top panel shows the selected on-pulse region. The pulsar at these frequencies is
    subject to scattering that gives rise to the exponential tail. \textit{Bottom:} The averaged 2~hr pulse profile.}
    \label{fig:0540}
\end{figure}

PSR~J0540--6919 is a young pulsar in the Large Magellanic Cloud that has been observed by the Parkes telescope to emit giant pulses \citep{2003ApJ...590L..95J} and possesses a twin-peaked average pulse profile (bottom panel Figure \ref{fig:0540}). In a test of the PTUSE filterbank mode, we observed this source for two hours on March 26 2019 and detected a large number of giant pulses. The brightest giant pulse has a peak flux density of 5.4\,Jy and a mean flux density of 92\,mJy and is shown in Figure\, \ref{fig:0540}. These flux values are estimated using the MeerKAT SEFD values. At its assumed distance \citep{2004MNRAS.355...31J} of 50\,kpc, this giant pulse would have a peak luminosity of 13500\,Jy\,kpc$^2$. If placed in M31, this giant pulse would have a peak flux of $\sim$25\,mJy and would be detectable by the FAST telescope.

\subsection{Nulling in PSR~J0633--2015}
\begin{figure}
    \centering
    \includegraphics[width=0.9\columnwidth]{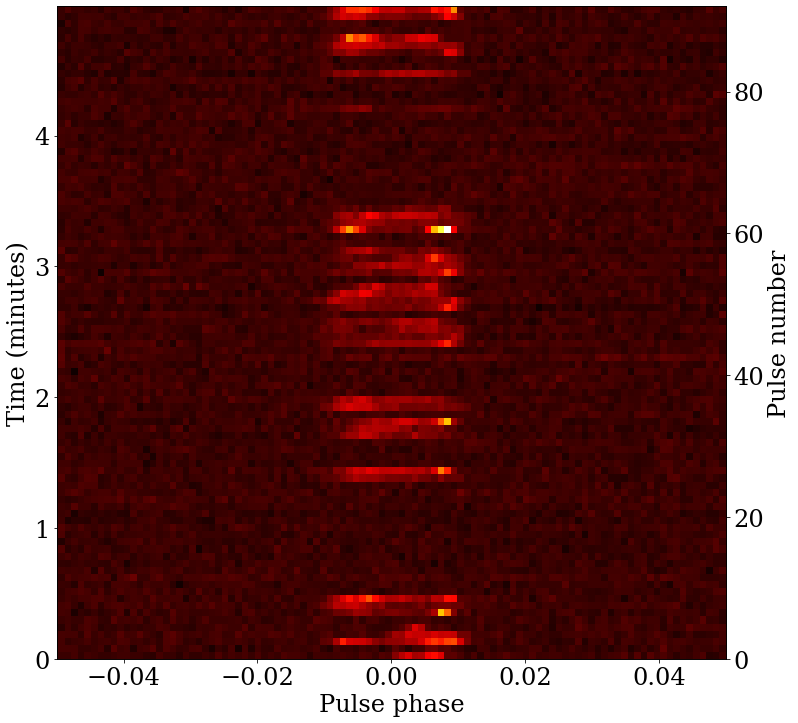}
    \caption{Short term nulling in PSR~J0633$-$2015 in observations made on 2019 October 27. Each horizontal row shows an individual pulse, with the colour coding denoting the flux density of the pulsar.}
    \label{fig:0633}
\end{figure}

The raw sensitivity of MeerKAT makes it ideal to study single pulse phenomenology in large number of pulsars
previously too weak for such studies. For example, understanding the population and characteristics of nulling pulsars, and how the pulsar sets the timescale for the on- and off-periods are key questions of the emission physics. PSR~J0633$-$2015 is a long period pulsar discovered by \citep{2006MNRAS.368..283B} who made no mention of its unusual pulse-pulse characteristics. Figure~\ref{fig:0633} shows a 5~minute observation made with MeerKAT on 2019 October 27. Each horizontal row shows a colour-coded representation of the flux density each individual pulse from the pulsar. Short duration nulling is clearly visible. 

\subsection{Double Pulsar Timing}

A full orbit of the double pulsar PSR~J0737$-$3039A \citep{2003Natur.426..531B,2004Sci...303.1153L} was observed to assess the precision of its arrival times with MeerKAT using 56 antennas. 
Using $16\times 53.5$\,MHz sub-bands and 64\,s integrations we obtained a post-fit rms
residual of just 9.3\,$\mu$s. Averaging across the full band
yields a post-fit arrival time rms of just 2.3\,$\mu$s in 64\,s.
By comparison, a 2013 archival observation of J0737$-$3039A with the Parkes 64\,m telescope using the multibeam receiver and the 340\,MHz CASPSR coherent dedisperser 64\,s integrations yielded 
15\,$\mu$s rms residuals. 
The improvement in timing is therefore a factor of 6.5.

We now examine how this compares with the radiometer equation expectations. The \textrm{SEFD} of the Parkes telescope/multibeam receiver is $\sim$ 30\,Jy whereas MeerKAT is $\sim$7\,Jy and the effective (RFI-free) bandwidths used are 340 and $\sim$756\,MHz respectively. Therefore, we would expect the ratio of the MeerKAT residuals to that of the Parkes multibeam CASPSR residuals to be of order $56/64 \times 7/30\times\sqrt{340/756}\sim 1/7$, 
in good agreement with our measured ratio of 6.5. Given the flux density variations exhibited by pulsars at the dispersion measure of this pulsar (48.9\,pc\,cm$^{-3}$) due to interstellar scintillation this is perfectly consistent with expectations based on our impressive telescope gain and system temperature figures quoted for MeerKAT's L-band receiver. A factor of 6.5 in timing residuals increases 
observing efficiency by a factor of $6.5^2 \sim 40$.

These test observations imply that MeerKAT will be important for studies of the double pulsar, in particular its eclipse, Shapiro delay, and hunts for the (now invisible) ``B'' pulsar that demand high sensitivity \citep[see e.g.][]{2004hpa..book.....L}. In the
longer term, extremely high precision will be required to separate the contributions of Lense-Thirring precession
to the relativistic advance of periastron and hence determine the moment of inertia of the neutron star in a novel way.

\section{Discussion and Outlook}
\label{sec:discussion}
The MeerKAT telescope was designed both to be a powerful standalone instrument and to be integrated into SKA1-mid. The development and commissioning of the PTUSE instrument has demonstrated that it is achieving excellent sensitivity, pulsar timing and polarimetric accuracy and already making discoveries such as new records on pulse jitter and timing stability. But MeerKAT has the potential to break other new ground. It is an extremely agile mechanical telescope, and can slew at 2 deg/s in azimuth and 1\,deg/s in elevation. The current dead time between pulsar observations is now just $\sim$5\,s 
leading to high observing efficiencies. In many regions of the galaxy, many pulsars will occupy the same primary beam, 
and in future array releases it will be possible to place a tied-array beam on up to four objects at once to further enhance the timing program efficiency. However other opportunities exist to further enhance pulsar timing at MeerKAT.
The TRAPUM and Breakthrough Listen compute clusters currently being commissioned on site 
receive the voltages from every individual antenna and can form coherent beams that can be independently steered within
the primary beam. TRAPUM will form up to (depending upon bandwidth) 900 coherent beams to search for
pulsars and Fast Radio Bursts \citep{sk16} whilst the Breakthrough Listen cluster intends to search for techno-signatures from advanced civilisations \citep{gajjar2019breakthrough}. 
These instruments can complement PTUSE by placing a tied beam on all known pulsars within the primary beam to maximise 
efficiency when timing the dozens of MSPs that inhabit globular clusters like 47 Tucanae and Terzan 5.

In February 2020 four completely independent sub-arrays were tested, each of them observing a different pulsar. It will soon be possible to time over 1000 pulsars in just an 8\,hour period using these modes.

All of the low dispersion measure MSPs ($DM<40$ pc cm$^{-3}$) exhibit scintillation maxima on the timescale of hours that
can easily amplify/deamplify their mean flux by factors of several. If we could observe MSPs during such maxima the benefits
in observing efficiency are significant. 
In the absence of limits from pulse jitter, a factor of just two in mean flux density is worth a four-fold improvement in timing efficiency according to the radiometer equation. Hence in the future 
we intend to use the bulk of the array to time a timing array MSP whilst a small group of antennas 
conduct an audit of potential targets that may be 
in a bright scintillation state. This could lead to dramatic increases in the sensitivity of the MeerKAT pulsar timing array
and its contribution to the IPTA's goal of detecting nanoHz gravitational waves.

The MeerTime Large Survey Project consists of four major sub-projects or themes: relativistic and binary pulsars, globular clusters, the MeerTime Pulsar Timing Array and the Thousand Pulsar Array. 
These all aim to create a legacy dataset for current and
future generations of astronomers. 
Data on the first three projects will be made available according to
SARAO Large Survey Project data release guidelines and upon
publication. The Thousand Pulsar Array \citep{johnstonetal2020} has an ambitious
objective to make its data public once it is cleaned, calibrated, 
and the timing corrections are secure.
As of Feb 26 2020 MeerTime\footnote{www.meertime.org} has already observed 1005 unique target pulsars in 825\,h of observing and shows that MeerKAT should
be an exceptional pulsar facility in the
lead-up to its incorporation into the SKA. Many
of the pointings were of globular clusters, and
hence well over 1000 individual pulsars have already obtained pulse
profiles suitable for timing and polarimetry.

\begin{acknowledgements}

Parts of this research were conducted by the Australian Research Council Centre of Excellence for 
Gravitational Wave Discovery (OzGrav), through project number CE170100004. FJ acknowledges funding from the European Research Council (ERC) under the European Union's Horizon 2020 research and innovation programme (grant agreement No. 694745). 
AS and JvL acknowledge funding from the Netherlands Organisation for Scientific Research (NWO) under project ''CleanMachine'' (614.001.301). YM, AS and JvL acknowledge funding from the ERC under the European Union's Seventh Framework Programme (FP/2007-2013) / ERC Grant Agreement n. 617199. A. Ridolfi gratefully acknowledges financial support by the research grant ``iPeska'' (P.I. Andrea Possenti) funded under the INAF national call PRIN-SKA/CTA approved with the Presidential Decree 70/2016. Pulsar  research  at  the Jodrell Bank Centre for Astrophysics is supported by a consolidated grant  from the Science and Technology Facilities Council (STFC).  The National Radio Astronomy Observatory is a facility of the National Science Foundation operated under cooperative agreement by Associated Universities, Inc.  SMR is a CIFAR Fellow and is supported by the NSF Physics Frontiers Center award 1430284.  Pulsar research at UBC is supported by an NSERC Discovery Grant and by the Canadian Institute for Advanced Research.
The MeerKAT telescope is operated by the South African Radio Astronomy Observatory, which is a facility of the National
Research Foundation, an agency of the Department of Science and Innovation. 
SARAO acknowledges the ongoing advice and calibration of GPS systems by the National Metrology Institute of South Africa (NMISA) and the time space reference systems department department of the Paris Observatory.
MeerTime data is housed on the
OzSTAR supercomputer at Swinburne University of Technology.
The MeerTime Pulsar Timing Array acknowledges support of the Gravitational Wave Data Centre funded by the Department of Education via Astronomy Australia Ltd. MK and DCJ acknowledge significant 
support from the Max-Planck Society (MPG). PTUSE was developed with support from 
the Australian SKA Office and Swinburne University of Technology.

\end{acknowledgements}

\begin{appendix}


\end{appendix}

\bibliographystyle{pasa-mnras}
\bibliography{refs}

\end{document}